\documentclass[epj,nopacs]{svjour}
\usepackage{amsmath}
\usepackage{amssymb}
\usepackage{latexsym}
\usepackage[dvips]{graphicx}
\usepackage[english]{babel}

\def\bge{\begin{equation}}
\def\ene{\end{equation}}
\def\bgea{\begin{eqnarray}}
\def\enea{\end{eqnarray}}
\def\nn{\nonumber}

\def\imnim{i}

\def\rxt{$\mathrm{R\chi T}$~}

\def\bge{\begin{equation}}
\def\ene{\end{equation}}
\def\bgea{\begin{eqnarray}}
\def\enea{\end{eqnarray}}
\def\nn{\nonumber}

\def\ls{\raise 1.5pt\hbox{$\,<\;$}\kern -10.5pt\lower3.5pt
          \hbox{$\sim$}\kern 1.5pt} %%% less or similar
\def\gs{\raise 1.5pt\hbox{$\,>\,$}\kern -9.5pt\lower3.5pt
          \hbox{$\sim$}\kern 1.5pt} %%% greater or similar

\usepackage{color}

\usepackage{ulem} % use: \sout{Striked out text}

\begin{document}
\sloppy
\title{$e^+e^-$ annihilation to $\pi^0 \pi^0 \gamma$ and $\pi^0 \eta \gamma$
as a source of information on scalar and vector mesons
}
\author{S.~Eidelman~\inst{1,2} , S.~Ivashyn~\inst{3,4}, A.~Korchin~\inst{3},
G.~Pancheri~\inst{5}
    and
     O.~Shekhovtsova \inst{3,5}
        }
\institute{
${}^1$
Budker Institute of Nuclear Physics, Novosibirsk, Russia
\\
${}^2$
Novosibirsk State University, Novosibirsk, Russia
\\
${}^3$
Institute for Theoretical Physics, NSC ``Kharkov Institute of
Physics and Technology'', UA-61108 Kharkov, Ukraine
\\
${}^4$
Institute of Physics, University of Silesia, PL-40007 Katowice, Poland
\\
${}^5$
INFN LNF, Frascati (RM) 00044, Italy
\\
\email{{simon.eidelman@cern.ch}, {ivashin.s@rambler.ru}, \\
        {korchin@kipt.kharkov.ua}, {giulia.pancheri@lnf.infn.it},
        {shekhovtsova@kipt.kharkov.ua} }
}

\authorrunning{S.~Eidelman {\it et al.}}
\titlerunning{$e^+e^-$ annihilation to $\pi^0 \pi^0 \gamma$
and $\pi^0 \eta \gamma$
as a source of information on scalar and vector mesons
}

%\date{\today}
%\date{March 10, 2010}
%\date{May 07, 2010}
\date{July 08, 2010}
%\date{}

\abstract{
%=====================================================================
We present a general framework for the model-independent decomposition of
the fully differential cross section of the reactions $e^+e^- \to
\gamma^\ast \to \pi^0\pi^0\gamma$ and $e^+e^- \to \gamma^\ast \to
\pi^0\eta \gamma$, which can provide important information on the
properties of scalar mesons: $f_0(600)$, $f_0(980)$ and $a_0(980)$.
For the model-dependent ingredients in the differential cross
section, an approach is developed, which relies on Resonance Chiral
Theory with vector and scalar mesons. Numerical results are compared to
data. The framework is convenient for development of a Monte Carlo
generator and can also be applied to the reaction $e^+e^- \to \gamma^\ast
\to \pi^+\pi^-\gamma$.
%=====================================================================
     % end of PACS codes
}
 %end of abstract
\maketitle

% ---------------------------------------------------------------------------------
\section{Introduction}
\label{intro}

Despite extensive studies during last decades,
physics of the light scalar mesons $a_0(980)$
($I^G(J^{PC}) = 1^-(0^{++})$), $f_0(980)$ and  $f_0(600) \equiv
\sigma$ ($I^G(J^{PC}) = 0^+(0^{++})$) is far from complete
understanding. In particular, there are doubts whether simple quark
model can explain their properties,
see, e.g., the review in~\cite{PDG_2008}.

The dominant decay channels of scalar mesons are known to be
$\pi^+ \pi^-$, $\pi^0 \pi^0$  for the $f_0 (980)$ and $\sigma$ meson,
and $\pi^0 \eta$ for the $a_0(980)$ meson.
Much experimental attention has already been paid  to the radiative
decays of the $\phi$ meson: $\phi(1020) \to \gamma a_0 \to \gamma \pi\eta$
~\cite{Aloisio:2002bsa,Ambrosino:2009py}
and $\phi(1020) \to \gamma f_0 \;(or \; \gamma \sigma)\to \gamma \pi\pi$
~\cite{KLOEres,KLOEres:07}
(see also the KLOE summary in~\cite{KLOE:2009:scalarsummary} and
results from Novosibirsk~\cite{CMD2res,SNDres,Achasov:2000ym}).
Such measurements are a good source of
information about the scalar meson properties~\cite{Achasov_Ivanchenko}.
Various models
have been proposed to describe these
decays,~\cite{Achasov_Ivanchenko,Close:1992ay,Ivashyn:2007yy,Oller:2002na,Bramon:2002iw},
to mention a few. The calculated decay widths turn out to be very
sensitive to model ingredients, however, the experimental data is
still insufficient to unambiguously discriminate between the
models.

In the case of the neutral final state (FS), i.e.,
$\pi^0\pi^0\gamma$ and $\pi^0\eta \gamma$, the cross section is
determined solely by final-state radiation (FSR) mechanism, since
there is no initial-state radiation (ISR) contribution resulting
in the same final state. Despite the lower value of the cross
section, compared to the charged pion case ($e^+e^-\to
\pi^+\pi^-\gamma$), processes with the neutral-meson FS are an
invaluable source of information on complicated hadron dynamics.

In this paper we describe the differential cross section
of the $e^+ e^-$ annihilation
to a pair of neutral pseudoscalar mesons and one photon in the FS,
 \bge
 e^+ (p_+) \; e^- (p_-) \to \gamma^\ast \to  P_1 (p_1) \; P_2 (p_2) \; \gamma (k).
 \label{eq:reaction_P1P2}
 \ene
The pseudoscalar mesons ($J^{PC} = 0^{-+}$)  are denoted by $P_1 P_2
\equiv \pi^0 \pi^0$ and $\pi^0 \eta$. In Section~\ref{fsr_model} we
present a formalism for a differential cross section, which is the main
task of this paper. We provide more general 
formulae in comparison with
Refs.~\cite{Dubinsky:2004xv,Isidori:2006we,Achasov:1999wr}, namely,
the non-integrated expressions are
given as well as those integrated over the angles. It gives a
convenient ground to implement the results in the Monte Carlo generators,
e.g., in FASTERD~\cite{Shekhovtsova:2009yn} (based on the general
structure given in Ref.~\cite{Dubinsky:2004xv}) or
PHOKHARA~\cite{Grzelinska:2008eb}.

Our framework is consistent with symmetries of the strong
and electromagnetic interactions.
It incorporates a model-dependent description of the FSR
only through the explicit form of the Lorentz-invariant functions
$f_{1,2,3}$ and has a model-independent tensor decomposition.

In Sections~\ref{section_scal} and~\ref{section_double} we calculate the
FS hadronic tensor. It is the second goal of the paper to provide
such a description in terms of functions $f_{1,2,3}$. Our model
relies on the Lagrangian of Resonance Chiral Theory
(\rxt)~\cite{EckerNP321}. The \rxt is a consistent extension of Chiral
Perturbation Theory to the region of energies near 1 GeV, which
introduces the explicit resonance fields and exploits the idea of
resonance saturation. One of the advantages of the \rxt Lagrangian at
leading order (LO), which makes it
convenient for the present study, is that, having a good predictive
power, it contains very few free parameters compared with other
phenomenological models. In order to get good agreement with data,
we release a rigor of \rxt and include some $SU(3)$ symmetry breaking
effects (e.g., use realistic masses of vector mesons) and mixing
phenomena (e.g., a G-parity-violating $\phi\omega\pi^0$ transition).

The loop contributions follow from the model Lagrangian. For
example, the kaon loop in the $\phi f_0 \gamma$ transtion, which is often
considered as a pure phenomenology manifestation, in the present model is a
direct consequence of the \rxt Lagrangian. In order to simplify the
formulae, some numerically irrelevant loop contributions are omitted. In
addition, the resonance exchanges in the loops are not considered to
avoid problems with renormalizability.

We consider in detail the following intermediate states with scalar
and vector resonances, which lead to the same FS $P_1 P_2 \gamma$:
\begin{eqnarray}
&& \text{ scalar decay, (Section~\ref{section_scal})}\nn \\
 e^+e^-&\to& \gamma^\ast \to
S\gamma\to P_1P_2\gamma
 \label{fsr_proc_scal}
\\
 e^+e^-&\to& \gamma^\ast \to V \to S\gamma\to P_1P_2\gamma
 \nn
\\
&& \text{ vector contribution, (Section~\ref{section_double})} \nn
\\
\label{fsr_proc_vec}
e^+e^-&\to& \gamma^\ast \to V P_{1,2}\to
P_1P_2\gamma
\\
\nn e^+e^-&\to& \gamma^\ast \to V_a \to V_b P_{1,2}\to P_1P_2\gamma
\end{eqnarray}
where $S$ ($J^{PC} = 0^{++}$) is an intermediate scalar meson
($S=f_0$, $\sigma$ for $\pi_0 \pi_0$ FS and $S=a_0$ for
$\pi_0\eta$).
Only the lowest nonet of vector mesons
($V, \ V_a, \ V_b =\rho$, $\omega$ and $\phi$) is taken into account.

We are interested in the center-of-mass energy $\sqrt{s}$ range
from the threshold up to $M_\phi$.
This framework may also be used in a somewhat dedicated
case of $\sqrt{s}= M_\phi$, giving, e.g., the $\phi$ radiative decay
description.

For the quantitative illustration of our approach,
in Section~\ref{section_numer} we show the
numerical results for the values of $\sqrt{s} = 1$~GeV and
$\sqrt{s} = M_\phi$.
The meson-pair invariant mass distributions are of interest,
and for $\sqrt{s} = M_\phi$ they are compared with available
results from KLOE.
We demonstrate the interplay of the
contributions~(\ref{fsr_proc_scal})
and~(\ref{fsr_proc_vec}).
Conclusions follow in Section~\ref{section_conlus}.

%=====================================================
%                           General structure
%=====================================================
\section{General structure of the FSR cross section}
\label{fsr_model}

\begin{figure}
\begin{center}
\resizebox{0.29\textwidth}{!}{%
  \includegraphics{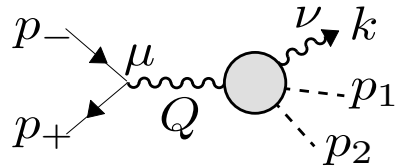}
  }
\end{center}
\caption{Generic scheme for electron-positron annihilation into two
particles with final state radiation
 }
\label{fig:e+e-generic-scheme}
\end{figure}

For a generic reaction $e^+ e^- \to \gamma P_1 P_2$ we define 4-momenta as
shown in Fig.~\ref{fig:e+e-generic-scheme}:
 \bgea
p&=& p_1 + p_2 , \quad \quad l = p_1 - p_2,  \\
Q&=& p_+ + p_- = k + p_1 + p_2 . \nn
 \enea
The masses of
pseudoscalars are $m(P_1)=m_1, \ m(P_2)=m_2$.

The cross section of the FSR  process can be written as
\begin{eqnarray}
\label{sect_fsr}
d\sigma_{F} &=& \frac{1}{2s(2\pi)^5}C_{12}
\nn\\&&\nn\!\!\!\!\!\!\!\!\!\!\!\!\times
\int
\delta^4(Q-p_1-p_2-k) \overline{|M_{FSR}|^2}
\frac{d^3p_1 \, d^3p_2\, d^3k}{8E_1E_2\,\omega}  \\
& = & C_{12} N\int \overline{|M_{FSR}|^2} d\cos\theta \, d\phi \, dm_{1\gamma}^2
\, dp^2 ,
\\\nn N& = &
\frac{1}{(2\pi)^4}\;\frac{1}{64s^2} , \nonumber
\end{eqnarray}
where $s=Q^2$, $\theta$ is the azimuthal angle, $\phi$ is the polar angle of the
photon and $m_{1\gamma}^2=(k+p_1)^2$. The factor
$C_{12}=1/2$ for $\pi^0 \pi^0$ in the final
state and $C_{12}=1$ for $\pi^0 \eta$. The matrix element
$M_{FSR}$ is
\begin{equation}
M_{FSR}=\frac{e}{s}M^{\mu\nu} \; \bar u(-p_+)\gamma_\mu
u(p_-)\epsilon^\ast_{\nu} ,
\end{equation}
where $e = \sqrt{4\pi \alpha} \approx \sqrt{4\pi /137}\approx 0.303$ and
the FSR tensor  $M^{\mu\nu}$ can be decomposed into three
gauge-invariant independent tensors:
\begin{eqnarray}
\label{eqn:fsr}
&&M^{\mu \nu }(Q,k,l)\equiv -ie^{2}(\tau_{1}^{\mu \nu }f_{1}
 +\tau_{2}^{\mu\nu}f_{2}+\tau _{3}^{\mu \nu }f_{3}) ,    \\
&&\tau _{1}^{\mu\nu}=k^{\mu }Q^{\nu }-g^{\mu \nu }k\cdot Q,
\nonumber \\
&&\tau _{2}^{\mu\nu}
=k\cdot l(l^{\mu }Q^{\nu }-g^{\mu \nu }Q\cdot l)
 +l^{\nu }(k^{\mu }Q\cdot l-l^{\mu }k \cdot Q), \;  \nonumber \\
&&
\tau _{3}^{\mu \nu }
=Q^{2}(g^{\mu \nu }k\cdot l-k^{\mu }l^{\nu})+Q^{\mu }(l^{\nu }k\cdot Q-Q^{\nu }k\cdot l)
\nonumber
\end{eqnarray}
with the Lorentz-invariant functions
\bge
f_i \equiv f_i (Q^2, k \cdot Q, k \cdot l),
\ene
$i=1,2,3$.
If $m_1 = m_2$, these tensors coincide with those of Ref.~\cite{Dubinsky:2004xv,Drechsel:1996ag}.
One may also find a similar approach in~\cite{Achasov:1999wr,EidelmanKuraev,ArbuzovKuraev}.
We emphasize that the decomposition~(\ref{eqn:fsr}) is
model independent; the model dependence is contained in an
explicit form of functions $f_i$ only.
Notice that the scalar products can be written in terms of
the invariant masses:
\bgea
k\cdot Q &=& (s-p^2)/2, \nn
\\
k\cdot l &=& m_{1\gamma}^2 - m_1^2 - k\cdot Q, \nn
\\
Q\cdot l &=& k\cdot l + s\delta/2
,
\enea
where $\delta\equiv {2(m_1^2-m_2^2)}/{s}$.

For the matrix element squared and averaged over the
$e^+e^-$ polarizations we obtain
\begin{eqnarray}
\label{aik}
\overline{|M_{FSR}|^{2}} &=&\frac{e^{6}}{s^{2}}\biggl[%
a_{11}|f_{1}|^{2}+2a_{12}\mathrm{Re}(f_{1}f_{2}^{\ast
})+a_{22}|f_{2}|^{2} \nonumber
\\
&&\!\!\!\!\!\!\!\!\!\!\!\!\!\!\!\!\!\!\!\!\!\!\!\!
+\; 2\; a_{23}\;{Re}(f_{2}f_{3}^{\ast
})+a_{33}|f_{3}|^{2}+2a_{13}{Re}(f_{1}f_{3}^{\ast })\biggr], \label{fsr}
\end{eqnarray}
with the coefficients
 \begin{equation}
a_{ik} \equiv (\frac{s}{2}g_{\mu\rho}-p_{+ \mu} p_{- \rho}-p_{+
\rho} p_{- \mu})\tau_i^{\mu\nu}\tau_k^{\rho\lambda} g_{\nu\lambda},
\end{equation}
equal to
\begin{eqnarray}
a_{11} &=&\frac{1}{4}s \left(t_{1}^{2}+t_{2}^{2} \right) , \nn\\
a_{22} &=&\frac{1}{8} \biggl[ sl^{4}(t_{1}+t_{2})^{2}+4l^{2}
\bigl(%
u_{1}{}^{2} \left( s^{2}+s(t_{1}+t_{2})+t_{2}^{2} \right)
\nn\\&&
+u_{2}{}^{2} \left( s^{2}+s(t_{1}+t_{2})+t_{1}^{2} \right)
\nonumber \\
&&+ 2u_{1}u_{2} \left( s^{2}+s(t_{1}+t_{2})-t_{1}t_{2} \right)
\bigr)
\nn\\&&
+8s(u_{1}^{2}+u_{2}^{2})(u_{1}+u_{2})^{2} \biggr] \nonumber \\
& - & \bigl( 4u_{1}^2+ 4u_{2}^2 +
l^2(2s+t_{1}+t_{2})\bigr) \frac{s^2(u_1+u_2)\delta}{4}
\nn\\&&
+
\bigl( l^2s+2u_{1}^2+ 2u_{2}^2 \bigr) \frac{s^3\delta^2}{8}
 , \nn\\
a_{33}&=&-\frac{s^{2}%
}{2} \bigl(  t_{1}t_{2}l^{2}+2(u_{1}+u_{2})(u_{2}t_{1}+u_{1}t_{2})
\nn\\&&
-\delta s (u_{2}t_{1}+u_{1}t_{2})\bigr)
,
\nn
\end{eqnarray}
\begin{eqnarray}
a_{12} &=&\frac{1}{8}\biggl[
sl^{2}(t_{1}+t_{2})^{2}+4u_{1}^{2}(s^{2}+st_{2}+t_{2}^{2})
\nn\\&&
+4u_{2}^{2}(s^{2}+st_{1}+t_{1}^{2})
+4u_{1}u_{2}(2s^{2}+s(t_{1}+t_{2})-2t_{1}t_{2})
\nn\\&&
+2s^2 \left( t_1u_2+t_2u_1+2s(u_1+u_2) \right) \delta+s^4\delta^2\biggr], \nn\\
a_{13} &=&\frac{s}{4} \biggl[%
(u_{1}+u_{2})(st_{1}+st_{2}+t_{1}t_{2})-u_{1}t_{2}^{2}-u_{2}t_{1}^{2}
\nn\\&&
-\frac{\delta}{2}(t_1+t_2)s^2 \biggr],
\nn \\
a_{23} &=&\frac{s}{4}\biggl[%
l^{2}(u_{1}t_{2}-u_{2}t_{1})(t_{1}-t_{2})-2s(u_{1}+u_{2})^{3}
\nn\\&&
+2(u_{1}+u_{2})(u_{1}-u_{2})(t_{2}u_{1}-u_{2}t_{1})
\nn
\\
&&+\delta s \left( u_1u_2(4s+t_{1}+t_{2})+u_{1}^2(2s-t_2)+u_{2}^2(2s-t_2)
\right)
\nn\\&&
-\frac{\delta^2}{2}s^3(u_{1}+u_{2})%
\biggr] ,
\label{aik_coeff}
%\label{eq:a-ik}
\end{eqnarray}
where
\begin{eqnarray}
\label{eq:scalars}
t_{1}&\equiv &
(p_{-}-k)^{2}-m^2_e=-2p_{-}\cdot k,
\nn\\
t_{2}&\equiv&
(p_{+}-k)^{2}-m^2_e=-2p_{+}\cdot k,  \nonumber \\
u_{1} &\equiv &l\cdot p_{-}, \;\; u_{2}\equiv l\cdot p_{+}.
\end{eqnarray}
For numerical calculations
the relation $l^2 = 2(m_1^2 + m_2^2) - p^2$ may
be useful.

The Eqs.~(\ref{sect_fsr}) and~(\ref{aik}),
with the explicit expressions~(\ref{aik_coeff})
and~(\ref{eq:scalars}),
fix the whole model-independent part of the
differential cross section.
It is worth illustrating a relation of these formulae to
the partial differential cross section.
Taking into account the corresponding factors and integrating the
coefficients $a_{ik}$ over the angular variables of the final-meson
phase space we have
\begin{eqnarray}
\label{eq:dsigma_dm2dp2}
\nonumber
\frac{d\sigma}{dm^2_{1\gamma}dp^2} &=&
\frac{\alpha^3 C_{12}}{32 s}
\left(A_{11}|f_{1}|^{2}+2A_{12}\mathrm{Re}(f_{1}f_{2}^{\ast
})+A_{22}|f_{2}|^{2} \right.
\\
&&\!\!\!\!\!\!\!\!\!\!\!\!\!\!\!\!\!\!\!\!\!\!\!\!
\left. +2A_{23}{Re}(f_{2}f_{3}^{\ast
})+A_{33}|f_{3}|^{2}+2A_{13}{Re}(f_{1}f_{3}^{\ast }) \right) ,
\end{eqnarray}
where
\begin{eqnarray}
A_{11}&=&\frac{4 x^2}{3} ,
\nn\\
A_{12} &=&
\frac{2s}{3} \bigl[ (x_1-x_2)^2+x^2(\sigma-1+x)-2\delta
(x_1-x_2)+\delta^2 \bigr] ,
\nn\\
A_{13}&=& -\frac{4s}{3}x(x_1-x_2-\delta)
\nn\\
 A_{23} &=& -\frac{2s^2}{3}(x_1-x_2)(\delta-x_1+x_2)^2 ,
\nn\\
A_{22}&=&\frac{s^2}{3} \bigl[ (x_1-x_2)^4+2(x_1-x_2)^2(1-x)(\sigma-1+x)
\nn\\&&
+2x^2(\sigma-1+x)^2
\nonumber \\
&&-2\delta(x_1-x_2) \left( (x_1-x_2)^2+(\sigma-1+x)(x_1+x_2) \right)
\nn\\&&
+\delta^2 \left( (x_1-x_2)^2+2(\sigma-1+x) \right) \bigr],
\nonumber \\
A_{33}&=& \frac{2s^2}{3} \bigl[
(x_1-x_2)^2(1+x)-x^2(\sigma-1+x)
\nn\\&&
+\delta(\delta-(2+x)(x_1-x_2)) \bigr]
%\nonumber
\label{eq_aik_integr}
,
\end{eqnarray}
and

\begin{eqnarray}
  &x& = \frac{s-p^2}{s}, \text{\hspace{0.2cm}} x_1=\frac{2E_1}{\sqrt{s}}=
\frac{p^2+m_{1\gamma}^2-m_2^2}{s} ,
\nn\\
 &x_2&=\frac{2E_2}{\sqrt{s}}=
\frac{s+m_2^2-m_{1\gamma}^2}{s} ,  \text{\hspace{0.2cm}}
\sigma=\frac{2(m_1^2+m_2^2)}{s} .
\end{eqnarray}

For the case $m_1=m_2$ Eq.~(\ref{aik_coeff}) reduces to Eq.~(17)
of Ref.~\cite{Dubinsky:2004xv}.
Also the results (\ref{eq:dsigma_dm2dp2}),~(\ref{eq_aik_integr}) coincide with Eqs.~(2.7),~(2.8)
of~\cite{Isidori:2006we}.
However, for an MC generator, the expressions~(\ref{sect_fsr}) and~(\ref{aik})
with coefficients $a_{ik}$ are more convenient than~(\ref{eq:dsigma_dm2dp2}).

Integrating Eq.~(\ref{eq:dsigma_dm2dp2}) over
$m_{1\gamma}^2$ one obtains the distribution of the invariant mass
$\sqrt{p^2}$ of two pseudoscalar mesons:
\bgea
\label{eq:dsigma_dp2}
\frac{d\sigma}{d \sqrt{p^2}} &=&
2\sqrt{p^2} \int_{(m_{1\gamma}^2)_{min}}^{(m_{1\gamma}^2)_{max}}
d m_{1\gamma} \left( \frac{d \sigma}{d m_{1\gamma}\; d p^2} \right)
.
\enea
The bounds of integration
over $m_{1\gamma}^2$ at the fixed value of $p^2$ are determined by
\bgea
\label{eq:m1gamma:limits}
(m_{1\gamma}^2)_{max/min} &=&
\frac{s(p^2\sigma+s\delta)}{4p^2}
\nn\\
&&+\frac{s-p^2}{2}
\Biggl(1\pm\sqrt{1-\frac{s\sigma}{p^2}+\frac{s^2\delta^2}{4p^4}}\Biggr)
.
\enea

At the $\phi$-meson peak ($s=M_\phi^2$) one can present the
results in terms of the branching ratio for the $\phi \to P_1 P_2
\gamma$ decay, which is related to the cross section as follows:
 \bgea
 \label{eq:phi-br}
 \frac{d B(\phi\to P_1 P_2\gamma)}{d\sqrt{p^2}}&=&\frac{M_\phi^2}{12\pi B(\phi\to
e^+e^-)}
 \nn\\&&\times
\frac{d \sigma (e^+e^- \to P_1 P_2 \gamma)}{d\sqrt{p^2}} , \enea
where the $\phi \to e^+ e^-$ branching ratio $B(\phi\to e^+e^-)$
is used. In the context of this paper, a calculation of this
branching ratio is useful for comparison of model predictions with
available data.

%=====================================================%
%                           Scalar contribution
%
%=====================================================
\section{Scalar contribution}\label{section_scal}

\begin{figure}
\begin{center}
\resizebox{0.29\textwidth}{!}{%
  \includegraphics{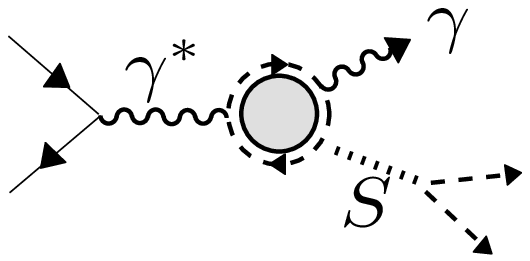}
  }
\end{center}
\caption{Scheme of $e^+ e^- \to S\gamma  \to P_1 P_2 \gamma$ subprocess
}
\label{fig:e+e-scalar-scheme}
\end{figure}

In this Section
we consider in detail the transition amplitudes
\begin{eqnarray} \label{fsr_proc_pi0pi0_vec}
&&\gamma^\ast\to f_0\gamma  \to \pi^0\pi^0 \gamma , \nonumber
\\
&&\gamma^\ast\to  \sigma \gamma  \to \pi^0\pi^0 \gamma , \nonumber
\\
\label{fsr_proc_pi0eta_vec}
&&\gamma^\ast\to a_0\gamma \to \pi^0\eta \gamma
\end{eqnarray}
for the $\pi^0\pi^0 \gamma$ and $\pi^0\eta \gamma$ final states,
respectively.
They contibute to $e^+ e^- \to S\gamma  \to P_1 P_2 \gamma$ as
illustrated in Fig.~\ref{fig:e+e-scalar-scheme}.
To describe the processes~(\ref{fsr_proc_pi0pi0_vec}) we use
the Lagrangian of \rxt~\cite{EckerNP321}
at the linear-in-resonance level, following~\cite{Ivashyn:2007yy,Ivashyn:2009te}.
%%%%%%
The basic features of the Lagrangian framework of the~\rxt
are sketched in~\ref{App:A}.
We emphasize that both light isoscalar scalar resonances, $f_0$ and $\sigma$
are included in the formalism in a natural way.
Throughout this section we work in the tensor representation for spin-$1$
particles~\cite{EckerNP321,EckerPLB223}.
In the present work we take into account
the pseudoscalar decay constants splitting ($f_\pi \neq f_K$)
which was discussed in the same context
in Ref.~\cite{Ivashyn:2009te}.

The interaction of pseudoscalars with the photon field $B^\mu$
in~\rxt is identical to the scalar QED.
We shall now discuss the interaction terms of the 
Lagrangian~(\ref{lagr:vec:master})
relevant to the processes~(\ref{fsr_proc_pi0pi0_vec})
(cf.~\cite{Ivashyn:2007yy}).
For the vector mesons in the even-intrinsic-parity sector one has
\begin{eqnarray}
\mathcal{L}_{\gamma V}
&=&
e F_V  F^{\mu \nu} \bigl(
\frac{1}{2}\rho^0_{\mu\nu} + \frac{1}{6}\omega_{\mu\nu} -
\frac{1}{3\sqrt{2}}\phi_{\mu\nu} \bigr),
\label{eq:F3}
\end{eqnarray}
\begin{eqnarray}
\label{eq:F4} \mathcal{L}_{VPP} & = & {i} G_V  \big[
\frac{1}{f_\pi^2}\;(2\ \rho^0_{\mu\nu}
\partial^\mu\pi^+
\partial^\nu \pi^- )
\nonumber
\\
&&\!\!\!\!\!\!\!\!\!\!\!\!\!\!\!\!
+ \frac{1}{f_K^2}
( \rho^0_{\mu\nu} +  \omega_{\mu\nu} - \sqrt{2}\phi_{\mu\nu} )
(\partial^\mu K^+\partial^\nu K^- )
\nonumber
\\
&&\!\!\!\!\!\!\!\!\!\!\!\!\!\!\!\!
+ \frac{1}{f_K^2}
(- \rho^0_{\mu\nu} +  \omega_{\mu\nu} - \sqrt{2}\phi_{\mu\nu} )
( \partial^\mu K^0\partial^\nu \bar{K}^0 )
\big],
\end{eqnarray}
\begin{eqnarray}
\label{eq:F5}
\mathcal{L}_{\gamma V PP} &=& -\frac{e F_V}{f_\pi^2}
\partial^\mu B^\nu \rho_{\mu \nu}^0  \ \pi^+ \pi^-
\nn
\\
&& -\frac{e F_V}{2 f_K^2}
\partial^\mu B^\nu \left(\rho_{\mu \nu}^0 + \omega_{\mu \nu} - \sqrt{2} \phi_{\mu
\nu}\right)\ K^+ K^- \nn
\\
&&- \frac{2e G_V}{f_\pi^2} B^\nu \rho_{\mu \nu}^0 \left(
\pi^+\partial^\mu \pi^-
 +  \pi^- \partial^\mu\pi^+\right)
\nn
\\
&&- \frac{e G_V}{f_K^2} B^\nu \left(\rho_{\mu \nu}^0 +
\omega_{\mu \nu} - \sqrt{2} \phi_{\mu \nu}\right)
 \nn\\&&\times
\left( K^+ \partial^\mu K^-  + K^-
\partial^\mu K^+ \right) , %\nonumber
\end{eqnarray}
where $F^{\alpha\beta}$ stands for the electromagnetic field
tensor and $V^{\mu \nu}$ for the vector field in the tensor
representation, $F_V$ and $G_V$ are the model parameters
(see~\ref{App_B} for numerical values). Vertex functions for
Eqs.~(\ref{eq:F3})--(\ref{eq:F5}) are shown in
Table~\ref{Table:v3}.

\begin{table*}
\caption{The vertices from Resonance Chiral Lagrangian
   terms~(\ref{eq:F3})-(\ref{eq:F5}).
   The dashed line stands for pseudoscalar meson (momentum~$l$),
   double solid --- for vector meson,
   wavy line --- for photon (momentum~$q$).}
\label{Table:v3}
\begin{center}
  \begin{tabular}
{|c|p{23pt}|p{23pt}|p{23pt}|p{35pt}|p{35pt}|p{35pt}|p{45pt}|p{45pt}|c|}
    \hline
       {Diagramm}  &
       \multicolumn{3}{c|}{\resizebox{0.17\textwidth}{!}{\includegraphics{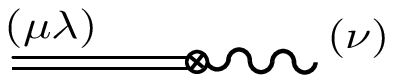}}} &
       \multicolumn{3}{c|}{\resizebox{0.15\textwidth}{!}{\includegraphics{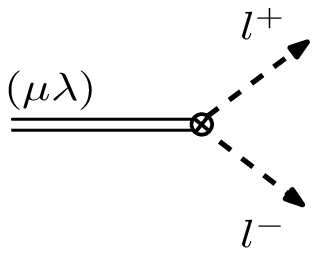}}} &
       \multicolumn{3}{c|}{\resizebox{0.15\textwidth}{!}{\includegraphics{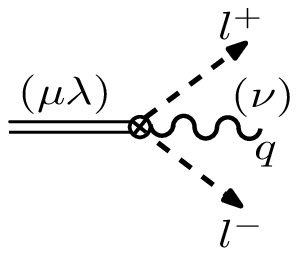}}} \\
%    \hline
       {Vertex function}  &
       \multicolumn{3}{c|}{$e F_V  \left[ g_{\nu\lambda}q_\mu - g_{\nu\mu} q_\lambda \right]$} &
       \multicolumn{3}{c|}{$\frac{G_V}{2 f_P^2} \left[ l^-_\mu l^+_\lambda - l^+_\mu l^-_\lambda\right]$} &
       \multicolumn{3}{c|}{$\frac{e G_V}{2 f_P^2} \left[ g_{\nu\lambda}(l^- + l^+)_\mu - g_{\nu\mu} (l^- + l^+)_\lambda \right]$} \\
       { }  &
       \multicolumn{3}{c|}{} &
       \multicolumn{3}{c|}{} &
       \multicolumn{3}{c|}{$+ \frac{e F_V}{4 f_P^2} \left[ g_{\nu\lambda}q_\mu - g_{\nu\mu} q_\lambda \right]$} \\
    \hline
       &
       \centering $\rho$ & \centering $\omega$ & \centering $\phi$ &
       \centering $\rho$ & \centering $\omega$ & \centering $\phi$ &
       \centering $\rho$ & \centering $\omega$ & $\phi$ \\
    \hline
       $\pi^\pm$ ($f_P = f_\pi$)
       &
       \multicolumn{3}{c|}{} &
       \centering$2$ & \centering$0$ & \centering$0$ &
       \centering$2$ & \centering$0$ &  $0$
\\
       $K^\pm$ ($f_P = f_K$)
       &
       \multicolumn{3}{c|}{} &
       \centering$1$ & \centering$1$ & \centering$-\sqrt{2}$ &
       \centering$1$ & \centering$1$ & $-\sqrt{2}$\\
%    \hline
       $K^0$ ($f_P = f_K$)
       &
       \multicolumn{3}{c|}{} &
       \centering$-1$ & \centering$1$ & \centering$-\sqrt{2}$ &
       \centering$0$ & \centering$0$ & $0$ \\
%    \hline
    \hline
%      Factor %due to $SU(3)$ flavor Clebsch-Gordan
&
      \centering$\frac{1}{2}$ & \centering$\frac{1}{6}$ & \centering$\frac{-1}{3 \sqrt{2}}$ &
      \multicolumn{6}{c|}{}\\
    \hline
  \end{tabular}
\end{center}
\end{table*}

\begin{figure}
\begin{center}
\resizebox{0.49\textwidth}{!}{%
   \includegraphics{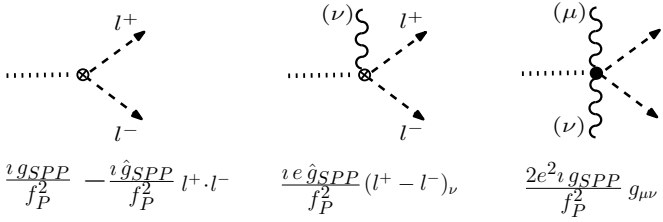}
   }
\end{center}
\caption{The vertices corresponding to
the Lagrangian~(\ref{eq:Lb}).
The dotted line stands for a scalar meson $S$,
the dashed one --- for a pseudoscalar~$P$.
Couplings are shown in
Table~\ref{table:generalscalarcouplings}.
} \label{fig:v1}
\end{figure}

The Lagrangian terms for scalar and pseudoscalar meson
interactions, which follow from~(\ref{lagr:master}) are
\bgea
\label{eq:Lb}
\nn
\mathcal{L}_{scalar}
&=&
\sum_{S} S \Bigl(
\frac{1}{f_\pi^2}\frac{g_{S\pi\pi}}{2}\stackrel{\rightarrow}{\pi}^2 +
\frac{1}{f_\pi^2}\frac{g_{S\eta\eta}}{2}\eta^2
+ \frac{1}{f_\pi^2}g_{S\pi\eta} \pi^0 \eta
\\&& %\!\!\!\!\!\!\!\!\!\!\!\!\!\!\!\!
\nn
+ \frac{1}{f_K^2}g_{SKK} \left(K^+K^- +(-1)^{I_S} K^0\bar{K}^0 \right)
\\&& %\!\!\!\!\!\!\!\!\!\!\!\!\!\!\!\!
+\frac{1}{f_\pi^2}(\hat{g}_{S\pi\pi}/2)(\partial_\mu\stackrel{\rightarrow}{\pi})^2
 \nn\\&&%\!\!\!\!\!\!\!\!\!\!\!\!\!\!\!\!
+ \frac{1}{f_\pi^2}(\hat{g}_{S\eta\eta}/2)(\partial_\mu\eta)^2
+ \frac{1}{f_\pi^2}\hat{g}_{S \pi^0 \eta}\partial_\mu\pi^0 \partial^\mu\eta
\nn\\&& %\!\!\!\!\!\!\!\!\!\!\!\!\!\!\!\!
+ \frac{1}{f_K^2}\hat{g}_{SKK} \left( \partial_\mu K^+\partial^\mu K^-
+(-1)^{I_S} \partial_\mu K^0 \partial^\mu \bar{K}^0 \right)
\nn\\&& %\!\!\!\!\!\!\!\!\!\!\!\!\!\!\!\!
+ \frac{1}{f_\pi^2}g_{S\gamma\pi\pi} eB_\mu \pi^+
\stackrel{\leftrightarrow}{\partial_\mu}\pi^-
\nn\\&& %\!\!\!\!\!\!\!\!\!\!\!\!\!\!\!\!
+ \frac{1}{f_K^2}g_{S\gamma KK}eB_\mu K^+ \stackrel{\leftrightarrow}{\partial_\mu}
K^-
\nn\\&& %\!\!\!\!\!\!\!\!\!\!\!\!\!\!\!\!
+ \frac{1}{f_\pi^2}g_{S\gamma\gamma\pi\pi}e^2B_\mu B^\mu \pi^+ \pi^-
\nn\\&& %\!\!\!\!\!\!\!\!\!\!\!\!\!\!\!\!
+ \frac{1}{f_K^2}g_{S\gamma\gamma KK}e^2B_\mu B^\mu K^+ K^- \Bigr).
     \enea
 (interactions with $\eta^\prime$ are omitted here for brevity).
     Here $S$ stands for any scalar field, $a_0$,$f_0$ or $\sigma$,
and $P$ -- for pseudoscalar $\stackrel{\rightarrow}{\pi}= \pi^0, \pi^\pm$ or
$K^\pm$, $K^0$, $\bar{K}^0$ and $\eta$.
We have introduced the effective couplings $g_{S \pi \pi }$,
$g_{S \eta\eta}$, etc. listed in
Table~\ref{table:generalscalarcouplings}, ${I_S}=0$ for $f_0$ and
$\sigma$ and ${I_S}=1$ for $a_0$.
Couplings are expressed in terms of the model parameters $c_d$, $c_m$
and $\theta$,
see also the expression~(\ref{eq:eta-coefficients}) for the $C_{q,s}$ coefficients.
The Lagrangian~(\ref{eq:Lb}) leads to the vertices shown in
Fig.~\ref{fig:v1}.

Given this set of interaction terms, the leading
contribution to the $\gamma^\ast \gamma S$ vertex comes from
the one-loop diagrams~\cite{Ivashyn:2007yy}.
The mechanism of the $\phi$ meson decay via the
 kaon loop was first considered in a different formalism
in~\cite{Achasov_Ivanchenko} and
is consistent with the data~\cite{KLOE:2009:scalarsummary}.
We would like to stress
that in the current approach the loop mechanism is a predicted subprocess
following directly from the Lagrangian, rather than an assumption.
%%%%
In particular, for the case of the $\pi^0 \pi^0 \gamma$
final state both the kaon and pion loops contribute.
The latter are very important in the region of the $\rho$ resonance
(recall that the $\gamma^\ast$ invariant mass
$\sqrt{s}$ is not constrained to the $\phi$ meson mass).

\begin{figure*}
\begin{center}

\resizebox{0.9\textwidth}{!}{%
  \includegraphics{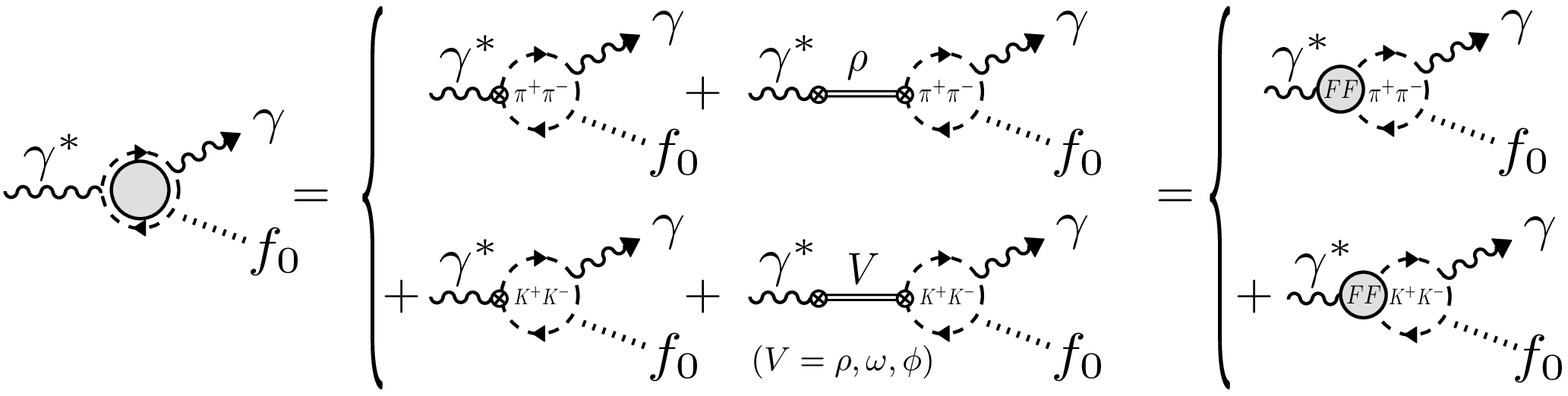}
  }

\vspace{25pt}

\resizebox{0.9\textwidth}{!}{%
  \includegraphics{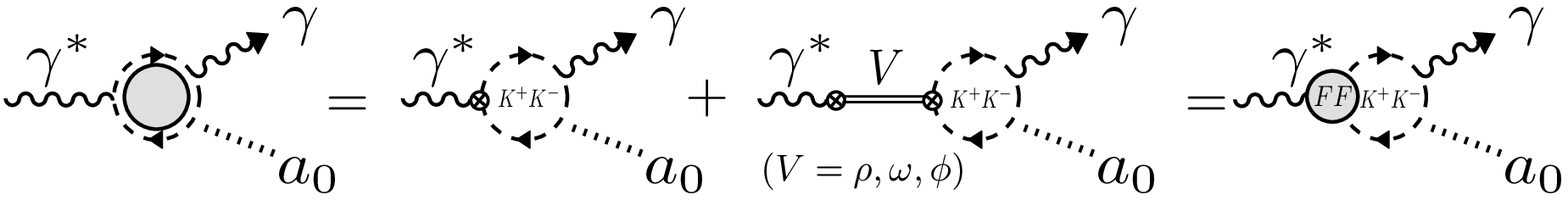}
}

\vspace{10pt}

\end{center}
\caption{Scheme for the $\gamma^\ast \gamma f_0$ and $\gamma^\ast \gamma \sigma$ (top) and
$\gamma^\ast \gamma a_0$ (bottom) transition.
Each ``loop blob'' corresponds to a set of diagrams following
from the Lagrangian, as explicitly shown in~\cite{Ivashyn:2007yy}
}
\label{fig:g-g-S-scheme}
\end{figure*}

When working with the three-point vertex functions $\gamma^\ast \gamma
S$, we factorize the kaon-loop part in the $a_0$ case
and separately the pion-loop and kaon-loop part for $f_0$ and $\sigma$,
as illustrated in Fig.~\ref{fig:g-g-S-scheme}
(see~\ref{App_C} for details).
The $\gamma^\ast (Q^\mu)\to \gamma(k^\nu) S (p)$ 
amplitude reads
 \bgea
 T^{\mu\nu} &=& -\imnim e^2 (Q^\nu k^\mu - g^{\mu\nu}Q\cdot k) 
 F_{S\gamma^\ast\gamma}(p^2,\!Q^2).
 \enea
The $\gamma^\ast (Q^2)\to \gamma S (p^2)$ transition 
form factors  (FF's) have the form
 \bgea
 F_{f_0\gamma^\ast\gamma}(p^2,\!Q^2) &=& G_{f_0\gamma^\ast\gamma}^{(\pi)}(p^2,\!Q^2) 
                                       + G_{f_0\gamma^\ast\gamma}^{(K)}(p^2,\!Q^2),
 \\
 F_{\sigma\gamma^\ast\gamma}(p^2,\!Q^2) &=& G_{\sigma\gamma^\ast\gamma}^{(\pi)}(p^2,\!Q^2) 
                                       + G_{\sigma\gamma^\ast\gamma}^{(K)}(p^2,\!Q^2),
 \\
 F_{a_0\gamma^\ast\gamma}(p^2,\!Q^2) &=& G_{a_0\gamma^\ast\gamma}^{(K)}(p^2,\!Q^2)
 ,
 \enea
where the terms
 \bgea
 \nn
\!\!\!\!G_{S\gamma^\ast\gamma}^{(\pi)}(p^2,\!Q^2)
&\!=&\! \frac{G_{S\pi\pi}(p^2)}{2\pi^2\; m_\pi^2}
I\!\left(\frac{Q^2}{m_\pi^2},\frac{p^2}{m_\pi^2}\right)\!F_{em}^{\pi} (Q^2),
 \\
\label{scalar-two-photon-ff-1}
\!\!\!\!G_{S\gamma^\ast\gamma}^{(K)}(p^2,\!Q^2)
&\!=& \!\frac{G_{S K K}(p^2)}{2\pi^2 \; m_K^2}
I\!\left(\frac{Q^2}{m_K^2}, \frac{p^2}{m_K^2}\right)\!F_{em}^{K} (Q^2),
 \enea
for $S=f_0,\sigma$, and
 \bgea
\label{scalar-two-photon-ff-3}
\!\!\!\!G_{a_0\gamma^\ast\gamma}^{(K)}(p^2,\!Q^2)
&\!=& \!\frac{G_{a_0 KK}(p^2)}{2\pi^2 \; m_K^2}
I\!\left(\frac{Q^2}{m_K^2}, \frac{p^2}{m_K^2}\right)\!F_{em}^{K} (Q^2)
 \enea
follow from~(\ref{eq:F3})--(\ref{eq:Lb}),
and the pion and kaon electromagnetic form factors,
$F_{em}^{\pi} (Q^2)$ and $F_{em}^{K} (Q^2)$,
follow from~(\ref{eq:F3}) and~(\ref{eq:F4}).
The terms
 \bgea
    G_{S KK}(p^2) &\equiv & 1/f_K^2 \left( \hat{g}_{S KK} (m_K^2\!-\!p^2/2)  + g_{S KK} \right), \nn\\
    G_{S \pi \pi}(p^2) &\equiv & 1/f_\pi^2 \left( \hat{g}_{S \pi \pi} (m_\pi^2 - p^2/2) + g_{S\pi \pi} \right),
\label{eq:SPP-ffs:1}
 \enea
for  $S= f_0,\sigma$ and
 \bgea
    G_{a_0 KK}(p^2) &\equiv & 1/f_K^2 \left( \hat{g}_{a_0KK} (m_K^2 - p^2/2) + g_{a_0KK} \right), \nn\\
    G_{a_0 \pi \eta}(p^2) &\equiv & 1/f_\pi^2 \left(
     \hat{g}_{a\pi\eta} (m_\eta^2 + m_\pi^2 - p^2)/2 + g_{a\pi\eta} \right)
\label{eq:SPP-ffs:2}
 \enea
have the meaning of momentum-dependent $SPP$ vertices.
The expression for $I(a,b)$ in~(\ref{scalar-two-photon-ff-1})--(\ref{scalar-two-photon-ff-3})
coincides with that of~\cite{Close:1992ay,Bramon:2002iw}
and for convenience is given in~\ref{App_C}.

The scalar meson contribution relevant to
the $\pi^0 \pi^0$ final state is
 \bgea
 \label{f1-scalar-f0}
 f_1^{S,\, \pi^0 \pi^0} &=& \sum_{S=f_0,\;\sigma}
D_{S}(p^2) G_{S\pi\pi}(p^2) \left(
G_{S\gamma^\ast\gamma}^{(\pi)}(p^2,Q^2) \right.
\nn\\&&
\quad \quad \quad\left. +
G_{S\gamma^\ast\gamma}^{(K)}(p^2,Q^2) \right),
\enea
and in the $\pi^0 \eta$ case one has
 \bgea
 \label{f1-scalar-a0}
\!\!\!\!f_1^{S,\, \pi^0 \eta} &=& D_{a_0}(p^2) G_{a_0 \pi\eta}(p^2)
G_{a_0 \gamma^\ast\gamma}^{(K)}(p^2,Q^2). %\nonumber
 \enea
We use the scalar meson propagator $D_S(p^2)$ in the
form~\cite{Ivashyn:2009te}
 \bgea
D_{S}^{-1}(p^2)&=& p^2 - M_S^2 +  M_S\; \Im\!\mathit{m}\!\left(
\tilde{\Gamma}_{S,\; {tot}}(M_S^2) \right)
\nn\\
&&+ i\, \sqrt{p^2}\;
\tilde{\Gamma}_{S,\; {tot}}(p^2)
 \enea
with
 \begin{eqnarray}
\tilde{\Gamma}_{tot,S}(p^2)&=&
\tilde{\Gamma}_{S\to\pi\pi}(p^2)%\theta(p^2-4m_\pi^2)
%\nn\\&&
+
\tilde{\Gamma}_{S\to K \bar{K}}(p^2)%\theta(p^2-4m_K^2)
, \ \ S=f_0,\sigma
\nonumber \\
\tilde{\Gamma}_{tot,a_0}(p^2)&=& \Gamma_{a_0\to \pi \eta}(p^2) %\theta(p^2-(m_\pi + m_\eta)^2)
%\nn\\&&
+ \tilde{\Gamma}_{a_0 \to K \bar{K}}(p^2)
.
%\theta(p^2-4m_K^2).%  \nonumber
\end{eqnarray}
Contributions of heavy particles to the total widths, e.g.,
$\Gamma_{f_0\to \eta\eta}(p^2)$, are neglected. Modified widths
$\tilde{\Gamma}$ in the above expressions are defined similarly to the
tree-level decay widths given in~\ref{App_B}, see Eqs.~(\ref{width:ape}),
but the analytic continuation is used: \bge \sqrt{f(p^2)} = e^{i\;
Arg(f(p^2))/2}\sqrt{|f(p^2)|} , \ene see Ref.~\cite{Ivashyn:2009te}.

By construction, the functions $f_1$ in~(\ref{f1-scalar-f0}),~(\ref{f1-scalar-a0})
are of the chiral order $\mathcal{O}(p^6)$: the diagrams of
Fig.~\ref{fig:g-g-S-scheme} are $\mathcal{O}(p^4)$ and
$SPP$ transition is $\mathcal{O}(p^2)$.

\begin{table}
\caption{Effective couplings for scalar mesons~\cite{Ivashyn:2009te}
    (to be used with vertices of Fig.~\ref{fig:v1}).
    Model parameters are $c_d$ and $c_m$; the
    scalar octet-singlet mixing angle $\theta$ is defined in Eq.~(\ref{eq:multiplet_sc});
    $\eta^\prime$ couplings are omitted; singlet couplings $\tilde{c}_d$ and $\tilde{c}_m$
    are related to $c_d$ and $c_m$ in the large-$N_c$ approximation.
    Notice that the entries relevant to the $\eta$ meson
    correct the results of Table 9 in Ref.~\cite{Ivashyn:2007yy}.
}
\label{table:generalscalarcouplings}
\begin{center}
\begin{tabular}{rcl}
\hline\noalign{\smallskip}
    $g_{f\pi\pi}$ &=& $- 2 \, c_m \, m_\pi^2 (2\, \cos \theta - \sqrt{2} \, \sin \theta)/\sqrt{3}$,
    \\
    $g_{f\eta\eta}$ &=& $- c_m (2\, (C_{s}^2(2 m_K^2 - m_\pi^2) + C_{q}^2 m_\pi^2) \, \cos \theta$
    \\&&
    $+ \sqrt{2}\, (C_{s}^2(4 m_K^2 - 2 m_\pi^2) - C_{q}^2 m_\pi^2) \, \sin \theta)/\sqrt{3}$,\\
    $g_{fKK}$ &=& $- c_m \, m_K^2(4 \, \cos \theta + \sqrt{2}\, \sin \theta)/\sqrt{3}$ .\\
\noalign{\smallskip}\hline\noalign{\smallskip}
    $\hat{g}_{f\pi\pi}$ &=& $2 \,c_d (2 \cos \theta - \sqrt{2}\, \sin \theta)/\sqrt{3}$,
    \\
    $\hat{g}_{f\eta\eta}$ &=& $c_d(2 (C_{q}^2 + C_{s}^2) \cos \theta$
    \\&&
      $- \sqrt{2}  (C_{q}^2 - 2 C_{s}^2) \sin \theta)/\sqrt{3} $ ,\\
    $\hat{g}_{fKK}$ &=& $c_d(4 \cos \theta + \sqrt{2} \sin \theta)/\sqrt{3}$.
\\
\noalign{\smallskip}\hline\noalign{\smallskip}
    $g_{{\sigma}\pi\pi}$ &=& $- 2 \, c_m \, m_\pi^2 (\sqrt{2}\, \cos \theta + 2 \, \sin \theta)/\sqrt{3}$,
    \\
    $g_{{\sigma}\eta\eta}$ &=& $- c_m (-\sqrt{2}\, (C_{s}^2(4 m_K^2 - 2 m_\pi^2) - C_{q}^2 m_\pi^2) \, \cos \theta$
    \\&&
    $+ 2\, (C_{s}^2(2 m_K^2 -  m_\pi^2) + C_{q}^2 m_\pi^2) \, \sin \theta)/\sqrt{3}$,\\
    $g_{{\sigma}KK}$ &=& $- c_m \, m_K^2( - \sqrt{2} \, \cos \theta + 4\, \sin \theta)/\sqrt{3}$ .\\
\noalign{\smallskip}\hline\noalign{\smallskip}
    $\hat{g}_{{\sigma}\pi\pi}$ &=& $2 \,c_d (\sqrt{2} \cos \theta + 2\, \sin \theta)/\sqrt{3}$,
    \\
    $\hat{g}_{{\sigma}\eta\eta}$ &=& $c_d(\sqrt{2} (C_{q}^2 - 2 C_{s}^2) \cos \theta$
    \\ &&
    $+ 2  (C_{q}^2 + C_{s}^2) \sin \theta)/\sqrt{3} $ ,\\
    $\hat{g}_{{\sigma}KK}$ &=& $c_d(- \sqrt{2} \, \cos \theta + 4\, \sin \theta)/\sqrt{3}$.
\\
\noalign{\smallskip}\hline\noalign{\smallskip}
    $g_{aKK}$            &=& $- \sqrt{2} \, c_m m_K^2$, \\
    $g_{a\pi\eta}$       &=& $-2 \sqrt{2}\, C_{q} \,c_m \, m_\pi^2$  ,\\
\noalign{\smallskip}\hline\noalign{\smallskip}
    $\hat{g}_{aKK}$      &=& $\sqrt{2} \, c_d$ ,\\
    $\hat{g}_{a\pi\eta}$ &=& $2 \sqrt{2} \, C_{q} \, c_d$
.\\
\noalign{\smallskip}\hline\noalign{\smallskip}
&&
      $g_{f\pi\eta}$ = $\hat{g}_{f\pi\eta}$ = $g_{\sigma\pi\eta}$ =
      $\hat{g}_{\sigma\pi\eta}$ = 0
,\\
&&
$g_{a\pi\pi}$  = $\hat{g}_{a\pi\pi}$  = $g_{a\eta\eta}$ = $\hat{g}_{a\eta\eta}$ = 0
.\\
\noalign{\smallskip}\hline
    $g_{S\gamma\pi\pi}$ &=& $- \imnim \hat{g}_{S\pi\pi}$, \hfill
    $g_{S\gamma KK}$ = $ - \imnim \hat{g}_{SKK}$ ,\nn\\
    $g_{S\gamma\gamma\pi\pi}$ &=& $\hat{g}_{S\pi\pi}$, \hfill
    $g_{S\gamma\gamma KK}$ = $\hat{g}_{SKK}$ \nn
\\
\noalign{\smallskip}\hline\noalign{\smallskip}
\end{tabular}
\end{center}
\end{table}

%*************************************************************
%                          VECTOR CONTRIBUTION
%*********************************************************
\section{Vector contribution}\label{section_double}

\begin{figure}
\begin{center}
\resizebox{0.45\textwidth}{!}{%
     \includegraphics{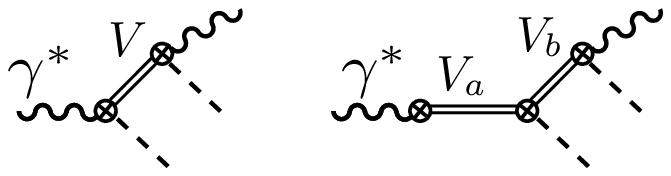}
      }
\end{center}
\caption{The vector, $\gamma^\ast \to V P_1 \to P_1 P_2 \gamma$,
and double vector, $\gamma^\ast \to V_a \to V_b P_1 \to P_1 P_2
\gamma$, contributions }\label{fig_vec}
\end{figure}

For $\gamma^\ast\to (\cdots)
\to\pi^0\pi^0\gamma$ the vector contribution mechanisms
are listed in Table~\ref{table:vector:contr:list} and the
corresponding diagrams are shown in Fig.~\ref{fig_vec}.

For the odd-intrinsic-parity vector-vector-pseudoscalar and
vector-photon-pseudoscalar interactions we use the chiral
Lagrangian in the vector formulation for spin-$1$ fields. As shown
in~\cite{EckerPLB237}, the use of vector formulation for $1^-$
fields ensures the correct behavior of Green functions to
order $\mathcal{O}(p^6)$, while the tensor formulation would
require additional local terms (see also discussion in
the Appendix~F of~\cite{Dubinsky:2004xv}). We choose Lagrangians of
Ref.~\cite{EckerPLB237,Prades}, that are $\mathcal{O}(p^2)$ and
$\mathcal{O}(p^3)$, for construction of the vector $\gamma V P $ and
double-vector $V V P$ contribution to $f_i$.
General Lagrangian terms are given in~\ref{App:A}.

Assuming exact $SU(3)$ case, the $\gamma V$
interaction can be written as
\begin{equation}
\mathcal{L}_{\gamma V} =  - e f_V  \partial^\mu B^\nu \bigl(
\tilde{\rho}^0_{\mu\nu} + \frac{1}{3}\tilde{\omega}_{\mu\nu} -
\frac{\sqrt{2}}{3}\tilde{\phi}_{\mu\nu} \bigr)
 \label{eq:vector_gamma_V}
\end{equation}
with $\tilde{V}_{\mu \nu} \equiv \partial_\mu V_\nu -
\partial_\nu V_\mu$ and $f_V = F_V / M_\rho$ is the coupling for
the vector representation of the spin-1 fields~\cite{EckerPLB223}.

The interactions of vector mesons in the odd-intrinsic-parity sector read
\begin{eqnarray}
\mathcal{L}_{V\gamma P}&=& -\frac{4\sqrt{2} e h_V}{3
f_\pi}\epsilon_{\mu\nu\alpha\beta} \partial^\alpha B^\beta \biggl[ (
\rho^{0\mu}  +3\omega^\mu + 3\varepsilon_{\omega\phi} \phi^\mu )
\partial^\nu \pi^0\nn \\ \label{lagr_vgp}
&&\!\!\!\!\!\!\!\!\!\!\!\!
 + \bigl[ (3 \rho^{0 \mu} + \omega^\mu)C_q + 2 \phi^\mu C_s \bigr] \partial^\nu \eta
 \biggr],
\end{eqnarray}
\begin{eqnarray}
\nn
\mathcal{L}_{VVP}&=&-\frac{4\sigma_V}{f_\pi}\epsilon_{\mu\nu\alpha\beta}
\biggl[
\pi^0 \partial^\mu \omega^\nu \partial^\alpha \rho^{0\beta}
\\
&&\!\!\!\!\!\!\!\!\!\!\!\!\!\!\!\!\!\!\!\!\!\!
+
\pi^0 \varepsilon_{\omega\phi}
\partial^\mu\phi^\nu \partial^\alpha \rho^{0\beta}
+
\pi^0 \varepsilon^\prime
\partial^\mu \omega^\nu \partial^\alpha \phi^{\beta}
\nn
\\
&&\!\!\!\!\!\!\!\!\!\!\!\!\!\!\!\!\!\!\!\!\!\!
 +  \eta \bigl[  (\partial^\mu\rho^{0\nu} \partial^\alpha
\rho^{0\beta}+
\partial^\mu \omega^{\nu} \partial^\alpha \omega^{\beta} )
  \frac{1}{2}\,C_q
  \nonumber \\
&&\!\!\!\!\!\!\!\!\!\!\!\!\!\!\!\!\!\!\!\!\!\!
 - \partial^\mu \phi^{\nu}\partial^\alpha \phi^{\beta}
\frac{1}{\sqrt{2}} \, C_s +
\varepsilon_{\omega\phi} \partial^\mu \phi^{\nu} \partial^\alpha
\omega^{\beta} (C_q + C_s)\bigr] \biggr],
\label{lagr_vvp}
\end{eqnarray}
where $\epsilon_{\mu \nu \alpha \beta}$ is the totally antisymmetric
Levi-Civita tensor.
As before, we omit the $\eta^\prime$ meson.

As it is also seen from (\ref{lagr_vgp}) and (\ref{lagr_vvp}), the
transitions $\gamma \phi \pi^0$, $\phi \rho^0 \pi^0$ and $\phi
\omega \eta$ are related to a small parameter
$\varepsilon_{\omega\phi}$, responsible for the  $u \bar{u} +
d\bar{d}$ component in the physical $\phi$ meson. The parameter
$\varepsilon^\prime$ is responsible for the G-parity-violating
$\phi\omega\pi^0$ vertex, caused by isospin breaking. The coupling
constants $f_V$, $h_V$ and $\theta_V$ are model parameters.
Numerical values for all parameters are given in~\ref{App_B}.

Due to a similar structure of the $\mathcal{L}_{V P \gamma}$ and $\mathcal{L}_{V V
P}$ interactions, the processes $\gamma^\ast\to V P_{1,2}\to P_1
P_2\gamma$ (one-vector-meson exchange) and $\gamma^\ast\to V_a
\to V_b P_{1,2} \to P_1 P_2 \gamma$ (double-vector-meson exchange)
can be described together. For this purpose it is convenient to
introduce the form factors $F_{\gamma^\ast V P}(Q^2)$ which
describe the transitions $\gamma^\ast (Q^2) \to V P$ including both
these mechanisms. Of course, the vector resonance enters
off-mass-shell.

For the $\gamma^\ast\to V  \pi^0 $ transition we obtain
\begin{eqnarray}
\label{eq:FFs_SU3_pi0}
F_{\gamma^\ast \rho \pi} (Q^2)&=& \frac{4}{3
f_\pi} \big[\sqrt{2} h_V - \sigma_V f_V {Q^2} D_\omega (Q^2)
\nn\\&&
 +
\varepsilon_{\omega\phi} \sqrt{2} \sigma_V f_V
 {Q^2} D_\phi (Q^2) \bigr],  \\
F_{\gamma^\ast \omega \pi}(Q^2)&=& \frac{4}{f_\pi} \bigl[\sqrt{2} h_V
- \sigma_V f_V {Q^2} D_\rho (Q^2)
\nn\\&&
 +
\varepsilon^\prime \frac{\sqrt{2}}{3} \sigma_V f_V {Q^2} D_\phi (Q^2)
\bigr], \nonumber \\
F_{\gamma^\ast \phi \pi}(Q^2) &=&  \varepsilon_{\omega\phi}
\frac{4}{f_\pi} \bigl[\sqrt{2} h_V
- \sigma_V f_V {Q^2} D_\rho (Q^2) \bigr]
. \nonumber
\end{eqnarray}
The vector meson $V= \rho, \omega, \phi$  propagators are
 \bgea
\label{vector-propagator-simple} D_V(Q^2) &= &[Q^2 - M_V^2 + \imnim
\sqrt{Q^2} \Gamma_{tot, V} (Q^2)]^{-1} .
 \enea
with an energy-dependent width for the $\rho$ meson
 \bgea
\Gamma_{tot, \rho}(Q^2) &=& \frac{G_V^2 M_\rho^2 }{48 \pi f_\pi ^4 Q^2}
\biggl[ \bigl(Q^2 - 4 m_\pi^2 \bigr)^{3/2}
\theta\bigl(Q^2 - 4 m_\pi^2 \bigr) \nn \\&&+ \frac{1}{2}
\bigl(Q^2 - 4 m_K^2 \bigr)^{3/2}\theta\bigl(Q^2 - 4 m_K^2 \bigr) \biggr]
 \label{eq:C1}
 \enea
and the constant widths for the $\omega$ and $\phi$ mesons.

In terms of these FF's we find the contribution to the functions $f_i$
(see Eq.~(\ref{eqn:fsr})) coming from the processes (\ref{fsr_proc_vec}).
For the $\pi^0 \pi^0 \gamma$ final state one obtains:
\begin{eqnarray}
\label{eq:delta-f1_pi0_pi0}
f_{1}^{V} &=& -\frac{1}{4} \sum_{V=\rho, \omega }
F_{\gamma^\ast V \pi} (Q^2)   F_{\gamma^\ast V \pi} (0)
\nn\\&&\!\!\!\!\!\!\!\!\!\!
\times
\bigl[ ({k\cdot Q +l^2}) \bigl(D_{V}(R^2_{+}) + D_{V}(R^2_{-})
\bigr)
\\
&& \nn + 2 k \cdot l
\bigl( D_{V}(R^2_{+}) - D_{V}(R^2_{-}) \bigr) \bigr] ,
\nonumber \\
f_{2}^{V} &= & \frac{1}{4} \sum_{V =\rho, \omega}
F_{\gamma^\ast V\pi}(Q^2)  F_{\gamma^\ast V \pi} (0)
\bigl[ D_{V} (R^2_{+})+ D_{V} (R^2_{-}) \bigr]
, \nonumber \\
f_{3}^{V} &=& -\frac{1}{4} \sum_{V =\rho, \omega}
F_{\gamma^\ast V \pi}(Q^2) F_{\gamma^\ast V \pi} (0)
\bigl[ D_{V}(R^2_{+}) - D_{V}(R^2_{-}) \bigr], \nn
\end{eqnarray}
where the contribution proportional to $F_{\gamma^\ast \phi \pi}(Q^2)
F_{\gamma^\ast \phi \pi}(0) \propto \varepsilon_{\omega\phi}^2$ has
been neglected. The momenta are defined as
    \bge
    R^2_{\pm} = (1/4) (Q^2 + l^2 +2 k\cdot Q \pm 2(k\cdot l+Q\cdot l) ),
    \ene
or equivalently $R^2_{+} = (k+ p_{1})^2$ and
$R^2_{-} = (k+ p_{2})^2$.

Similarly, for the $\gamma^\ast\to V   \eta $ transition we obtain FF's
\begin{eqnarray}
 \label{eq:FFs_SU3_eta}
 F_{\gamma^\ast \rho \eta}(Q^2) & =
 & C_q
F_{\gamma^\ast \omega \pi}(Q^2),  \\
 F_{\gamma^\ast \omega \eta}(Q^2) & = &
 C_q
F_{\gamma^\ast \rho \pi}(Q^2), \nonumber \\
 F_{\gamma^\ast \phi \eta}(Q^2) &=&
2\;C_s\; \frac{4}{3 f_\pi} \big[ \sqrt{2}
h_V - \sigma_V f_V {Q^2} D_\phi (Q^2) \big] \nonumber\\
 && -  \varepsilon_{\omega\phi}  (C_q+C_s)  \frac{4}{3 f_\pi} \sigma_V f_V
 {Q^2} D_\omega (Q^2) .
\nn
\end{eqnarray}

Correspondingly, the contribution to the functions $f_i$ for the $\pi^0
\eta \gamma$ final state is
\begin{eqnarray}
 \label{eq:delta-f1_pi0_eta}
f_{1}^{V} &=& -\frac{1}{4} \sum_{V=\rho, \omega, \phi }
\Bigl\{
F_{\gamma^\ast V \pi} (0) F_{\gamma^\ast V \eta} (Q^2)
\nn\\&&\times
 \bigl[ ({k\cdot Q
+l^2}) D_{V}(R^2_{+}) +
2 k \cdot l D_{V}(R^2_{+}) \bigr] \nonumber \\
&& + F_{\gamma^\ast V \eta} (0) F_{\gamma^\ast V \pi} (Q^2)
\nn\\&&\times
\bigl[ ( k\cdot Q +l^2 )
D_V (R^2_{-}) - 2 k \cdot l D_V (R^2_{-}) \bigr] \Bigr\},
    \nonumber \\
f_{2}^{V} &= & \frac{1}{4} \sum_{V=\rho, \omega, \phi } \Bigl\{
F_{\gamma^\ast V \pi} (0) F_{\gamma^\ast V \eta}(Q^2)  D_{V} (R^2_{+}) \Bigr.
\nn\\&&\Bigl.
+ F_{\gamma^\ast V \eta} (0) F_{\gamma^\ast V \pi}(Q^2)
D_{V} (R^2_{-}) \Bigr\},
\nonumber \\
f_{3}^{V} &= & -\frac{1}{4} \sum_{V=\rho, \omega, \phi } \Bigl\{
F_{\gamma^\ast V \pi} (0) F_{\gamma^\ast V \eta}(Q^2)  D_{V} (R^2_{+})  \Bigr.
\nn\\&&\Bigl.
-
F_{\gamma^\ast V \eta} (0) F_{\gamma^\ast V \pi}(Q^2) D_{V} (R^2_{-})
\Bigr\}.
\end{eqnarray}

%=====================================================%
%                           NUMERICAL
%
%=====================================================
\section{Numerical results}
\label{section_numer}
In this section we present the numerical
results obtained in our framework.
The model-dependent ingredients, namely, the functions
$f_{1,2,3}$ are given in Sections~\ref{section_scal}
and~\ref{section_double}.

The values of the model parameters, which we used in our numerical
results, are listed in~\ref{App_B}. The masses of vector
and pseudoscalar mesons are taken from~\cite{PDG_2008}. The
coupling of vector mesons to a pseudoscalar and photon $h_V$ is
estimated from the tree-level decay width. The scalar meson
couplings and mass parameters were found from the
fit~\cite{Ivashyn:2009te}.

\subsection{Scalar mesons and $\phi$ radiative decay}
\begin{figure}
\begin{center}
\resizebox{0.45\textwidth}{!}{%
     \includegraphics{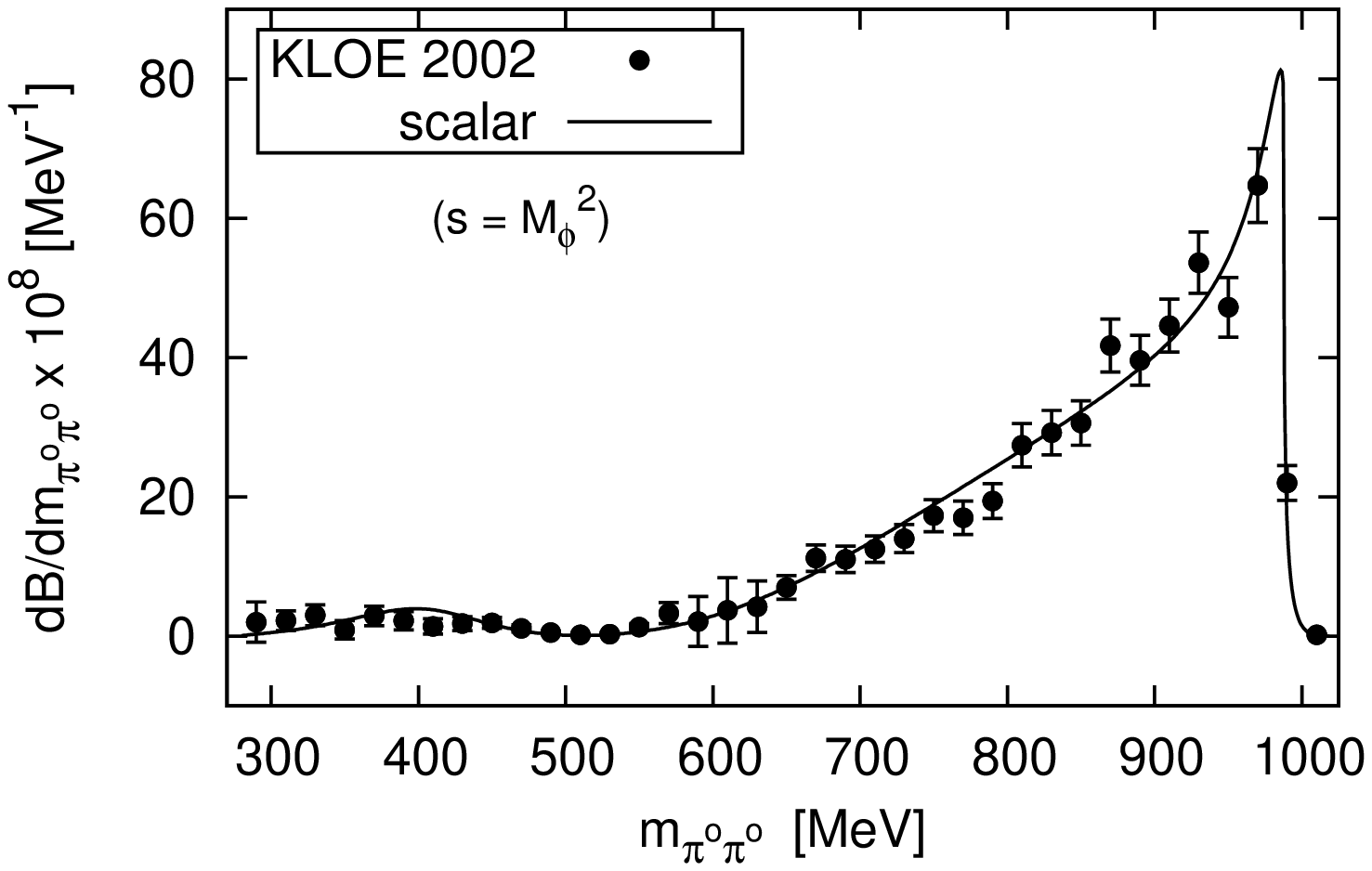}
      }

\resizebox{0.45\textwidth}{!}{%
     \includegraphics{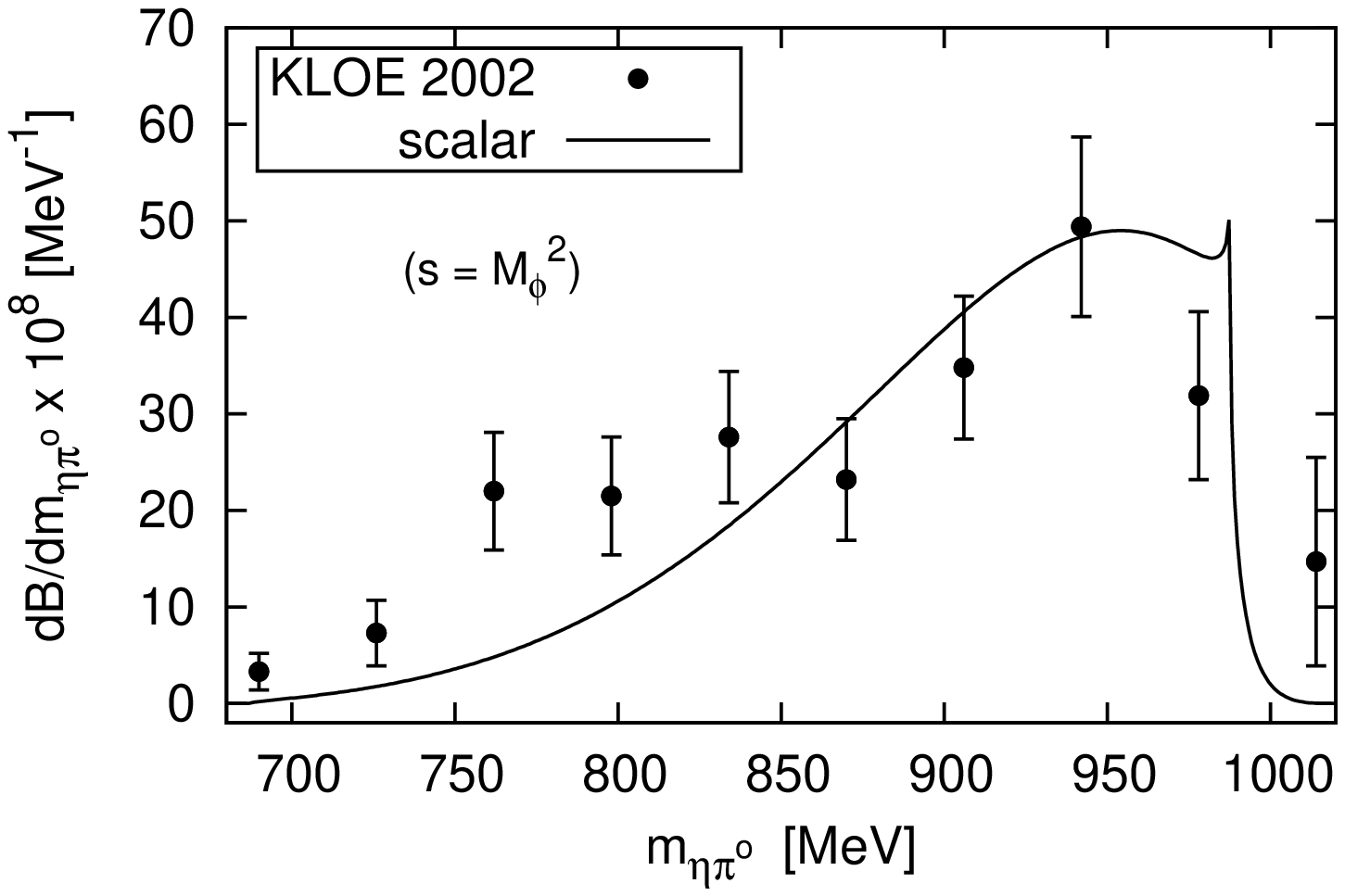}
      }

\resizebox{0.45\textwidth}{!}{%
     \includegraphics{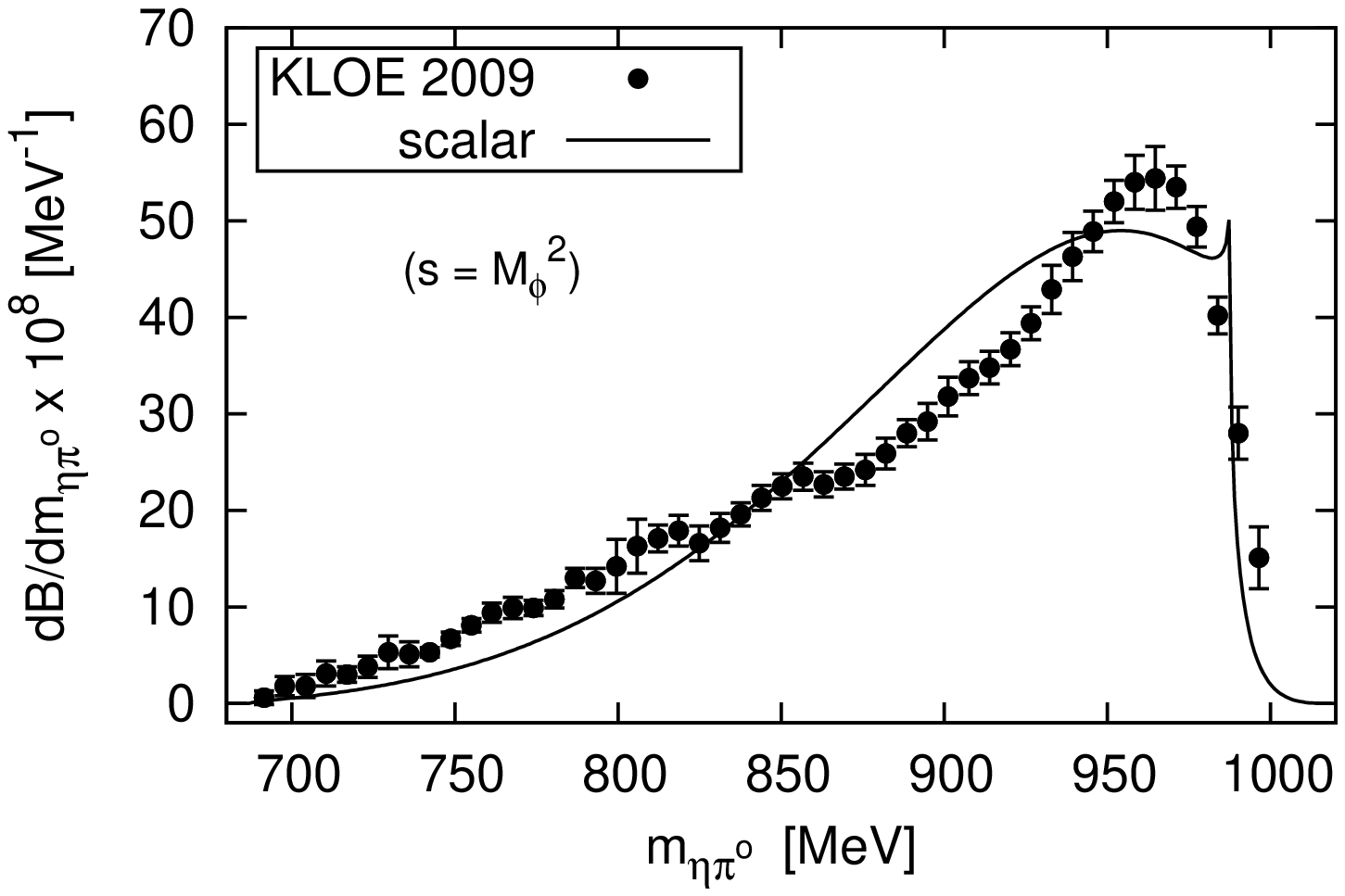}
      }
\end{center}
\caption{Invariant mass distributions
in the $e^+e^-$ annihilation
to $\pi^0 \pi^0 \gamma$ (top panel)
and $\pi^0 \eta \gamma$ (middle and bottom panel)
for $\sqrt{s}=M_\phi$.
Data are from~\cite{KLOEres} (top),~\cite{Aloisio:2002bsa} (middle)
and~\cite{Ambrosino:2009py} (bottom)
}
\label{fig:num:dBdm}
\end{figure}

As we discussed in this paper, in $e^+e^-$ annihilation
to $\pi^0 \pi^0 \gamma$ and $\pi^0 \eta \gamma$ both
scalar~(\ref{fsr_proc_scal}) and vector decays~(\ref{fsr_proc_vec})
contribute to the observed events.
The KLOE Collaboration has reported data
 on the invariant mass distributions~\cite{Aloisio:2002bsa,KLOEres}
at $\sqrt{s} = M_\phi$, in which the vector  meson contribution
has been subtracted. In~\cite{Ivashyn:2009te} we performed a
combined fit of ${d}B(\phi\to a_0 \gamma \to\pi^0\eta\gamma)/{d
\sqrt{p^2}}$ and ${d}B(\phi\to (f_0,
\sigma)\gamma\to\pi^0\pi^0\gamma)/{d \sqrt{p^2}}$ to the KLOE 2002
data~\cite{Aloisio:2002bsa,KLOEres}, considering only scalar meson
contributions. We have found the inclusion of the $\sigma$ meson
into the framework important, and have fixed the numerical values
of scalar meson couplings and 
mass parameters within the model, for
more detail see~\cite{Ivashyn:2009te}. In Fig.~\ref{fig:num:dBdm}
we show our model results for ${d}B(\phi\to S\gamma\to
P_1P_2\gamma)/{d \sqrt{p^2}}$, eq.~(\ref{eq:phi-br}), at $\sqrt{s}
= M_\phi$. In this and subsequent plots we use the notation
$m_{\pi^0\pi^0}$ and $m_{\eta\pi^0}$ for $\sqrt{p^2}$. Note that
only the scalar meson contribution to the $P_1P_2\gamma$ final
state is plotted in this Figure. The plot for the
$\pi^0\pi^0\gamma$ final state shows a rather good
fit~\cite{Ivashyn:2009te} to the KLOE 2002 data~\cite{KLOEres},
where both $f_0$ and $\sigma$ are taken into account.

\begin{figure}
\begin{center}
\resizebox{0.45\textwidth}{!}{%
     \includegraphics{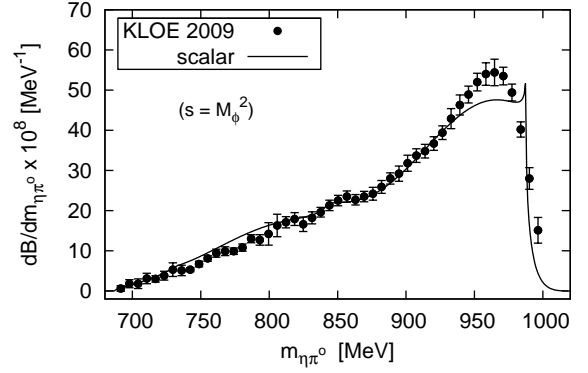}
      }
\end{center}
\caption{Invariant mass distributions
in the $e^+e^-$ annihilation to
$\pi^0 \eta \gamma$ for $\sqrt{s}=M_\phi$,
where the total contribution (vector and scalar)
is taken into account (cf.~Fig.~\ref{fig:num:dBdm} (bottom)).
Data are from~\cite{Ambrosino:2009py}.
}\label{fig:pieta:tot}
\end{figure}

In 2009 the new KLOE data~\cite{Ambrosino:2009py} on the
$\pi^0\eta\gamma$ channel appeared. A comparison of the model
prediction for $\phi\to a_0\gamma\to\pi^0\eta\gamma$ with these
new data is also shown in Fig.~\ref{fig:num:dBdm} (bottom). We
leave a refined fit of these new data for the future. Notice, if
one adds vector contributions to $\sigma(e^+e^-\to
\eta\pi^0\gamma)$ according to
Table~\ref{table:vector:contr:list}, then the shape of the
invariant mass distribution, calculated from
eq.~(\ref{eq:phi-br}), changes: cf.~Fig.~\ref{fig:num:dBdm}
(bottom) and Fig.~\ref{fig:pieta:tot}. It turns out that the 2009
KLOE data~\cite{Ambrosino:2009py} are better described by the
total contribution rather than by the scalar part alone. Note that
in Refs.~\cite{Ambrosino:2009py,SNDres} it was claimed that the
$\phi\to\pi^0\eta\gamma$ decay is dominated by the $\phi\to
a_0\gamma$ mechanism and the vector contribution is very small:
$B(e^+e^-\to VP \to \eta\pi^0\gamma) \ls 10^{-6}$.

% -----------------------------------------------
\begin{table}
\caption{Mechanisms of the vector contribution. Notice that some of
the channels, suppressed due to small parameters, can be enhanced in the
vicinity of the corresponding resonance (e.g.
$\gamma^\ast\to\phi\to\omega\pi^0$, see the text)}
\label{table:vector:contr:list}
\begin{center}
\begin{tabular}{|c|c|c|}
\hline & Dominant & Suppressed
\\
\hline \noalign{\smallskip} \noalign{\smallskip}
\multicolumn{3}{l}{in $\gamma^\ast\to (\cdots) \to\pi^0\pi^0\gamma$
:}
\\
\noalign{\smallskip} \hline 1-vector & $(\rho^0 \pi^0)$,
$(\omega \pi^0)$ & $(\phi\pi^0)$
\\
\hline 2-vector & \!$(\omega\!\to\!\rho^0\!\pi^0)$,
\!$(\rho^0\!\to\!\omega\!\pi^0)$ & $(\phi \to\rho^0\pi^0)$,
\!$(\phi\!\to\!\omega\!\pi^0)$
\\
&&$(\rho^0\!\to\!\phi\!\pi^0)$
%$(\rho^0\to\phi\pi^0)$ %% <- e^2
\\
\hline \noalign{\smallskip} \noalign{\smallskip}
\multicolumn{3}{l}{in $\gamma^\ast\to (\cdots) \to\pi^0\eta\gamma$ :
}
\\
\noalign{\smallskip}\hline 1-vector &
$(\rho\pi^0)$, $(\omega\pi^0)$ & $(\phi\pi^0)$
\\
& $(\rho\eta)$, $(\omega\eta)$ & $(\phi\eta)$
\\
\hline 2-vector & $(\rho\to\omega\pi^0)$,
$(\omega\to\rho\pi^0)$ & $(\rho\to\phi\pi^0)$,
$(\phi\to\rho\pi^0)$
\\
&
$(\rho\to\rho\eta)$, $(\omega\to\omega\eta)$
& $(\phi\to\phi\eta)$, $(\phi\to\omega\eta)$
%\\
%&&$(\phi\!\to\!\rho\eta)$
%$(\omega\to\phi\eta)$ %% <- e^2
\\
\hline
\end{tabular}
\end{center}
\end{table}

%%%%%%%%%%%%%%%%%%%%%%%%%%%%%%%%%%%%%%%%%%%%%%%%%%%%%%

\subsection{The $\gamma^* \to \rho \to\omega\pi$ and
$\gamma^*\to \phi  \to\omega\pi$ contribution}

\begin{figure}
\begin{center}
\resizebox{0.45\textwidth}{!}{%
     \includegraphics{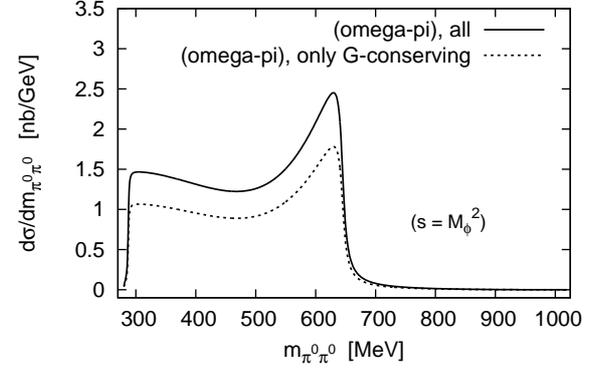}
      }
\end{center}
\caption{Partial differential cross section
of $e^+e^-$ annihilation to
$\pi^0 \pi^0 \gamma$ for $\sqrt{s}=M_\phi$ due to the
$\gamma^*\to\rho \to\omega\pi$ mechanism compared to
$\gamma^*\to(\rho,\ \phi)\to\omega\pi$
}\label{fig:pipi:epsprime}
\end{figure}

For the moment, to follow KLOE analysis~\cite{KLOEres:07} we neglect the
$G$-parity-violating vertex $\phi \omega \pi^0$, i.e.,
we set $\varepsilon^\prime = 0$.
For illustration we introduce the constant
$C_{\omega\pi}^\rho$~\cite{KLOEres:07,Shekhovtsova:2009yn}.
This constant can be obtained in terms of form
factors~(\ref{eq:FFs_SU3_pi0})
 \bgea
\frac{C^\rho_{\omega\pi}(s)}{16 \pi \alpha} & = & -\frac{1}{4}\;
F_{\gamma^\ast \omega \pi}(s)\; F_{\gamma^\ast \omega \pi}(0),
 \enea
leading to
 \bgea C^\rho_{\omega\pi} & = & -16\pi \alpha \;
\frac{4\sqrt{2} h_V}{f_\pi^2}\; \left( \sqrt{2} h_V - \sigma_V f_V s
D_\rho(s) \right)
\nn\\
& \approx & (0.597 - 0.542\;i) \ \text{GeV}^{-2}
 \enea
at $\sqrt{s}=M_\phi$. The KLOE result~\cite{KLOEres:07} for the same
constant is \
$C^\rho_{\omega\pi} = 0.850$ GeV$^{-2}$ ($\sqrt{s}=M_\phi$). Thus our
prediction for the absolute value, $\left|C^\rho_{\omega\pi}\right| =
0.751$ GeV$^{-2}$, which includes only the $\gamma^*(\to\rho)\to\omega\pi$
mechanism, is smaller than that of KLOE by about $15\%$.

This difference can be attributed to the $\rho^\prime=\rho(1450)$ meson which
is not included in the present calculation. To estimate the role of the
$\rho^\prime$ in the constant $C^\rho_{\omega\pi}$, we follow
Ref.~\cite{Dumm:2009va} (Eqs.~(32), (33)):
 \bgea
C^\rho_{\omega\pi} & = & -16\pi \alpha
\; \frac{4\sqrt{2} h_V}{f_\pi^2}\; ( \sqrt{2} h_V  \\
&-& \sigma_V f_V \frac{s}{1+\beta_{\rho^\prime}} (D_\rho(s)+\beta_{\rho^\prime}
D_{\rho^\prime}(s)) )
 \nn \\
& \approx & (1.06 - 0.69\;i) \ \text{GeV}^{-2} \nn
 \enea
for $\beta_{\rho^\prime}=-0.25$, $M_{\rho^\prime}=1.465$ GeV,
$\Gamma_{\rho^\prime}(M^2_{\rho^\prime})=400$ MeV and obtain
$\left|C^\rho_{\omega\pi}\right| = 1.27$ GeV$^{-2}$.

Next we turn on the parameter $\varepsilon^\prime$ responsible for the
$G$-parity-violating $\phi\pi\omega$ vertex and check how the $
C^\rho_{\omega\pi}$ value  changes.  Omitting $\rho^\prime$ we have
 \bgea
C^\rho_{\omega\pi} & = & -16\pi \alpha \; \frac{4\sqrt{2} h_V}{f_\pi^2}\;
( \sqrt{2} h_V - \sigma_V f_V s D_\rho(s)
\\ & + &\frac{\sqrt{2}}{3}\sigma_V f_V s \varepsilon^\prime D_\phi(s) ) \approx
(0.52 - 0.72\;i) \ \text{GeV}^{-2} \nonumber
 \enea
and obtain $\left|C^\rho_{\omega\pi}\right| = 0.892$ GeV$^{-2}$. While
making this estimation the value $\varepsilon^\prime = -0.0026$ has been
chosen~\footnote{Of course the experimental decay width
$\phi\to\omega\pi$ determines only the absolute value of this
parameter.}. Apparently, the present model with the lowest nonet of
vector mesons, supplemented with the $G$-parity-violating effect, allows
one to obtain the value for $C^\rho_{\omega\pi}$ close to the KLOE value
0.850 GeV$^{-2}$. Influence of the $\varepsilon^\prime$ parameter
on the cross section is presented in Fig.~\ref{fig:pipi:epsprime}.

Therefore, the difference between the $ C^\rho_{\omega\pi}$ value
originating from the $\gamma^*(\to\rho)\to\omega\pi$ mechanism,
and the value measured by KLOE may be explained by the
$\rho^\prime$ meson and/or $G$-parity-violating contribution. To
clarify further this issue, an analysis of data at $s=1 \
{\text{GeV}}^2$ will be essential~\footnote{At $s=1 \ {\text
{GeV}}^2$ the $G$-parity-violating vertex is suppressed, whereas
the $\rho^\prime$ mechanism survives. Therefore, any difference in
the values of $C^\rho_{\omega\pi}$ at two energies, $s=1 \ {\text
{GeV}}^2$ and $s=M_\phi^2$, would indicate sizeable
$G$-parity-violating effects.}.

\subsection{The $\gamma^* \to \phi \to\rho\pi$
 and $\gamma^*\to \omega \to\rho\pi$}

In a similar manner one can define $C_{\rho\pi}(s)$:
 \bgea
 \label{eq:Crhopi}
- 16\pi\alpha \frac{1}{4}\; F_{\gamma^\ast \rho \pi}(s)\; F_{\gamma^\ast
\rho \pi}(0) &=&
C_{\rho\pi}(s)
\\
 &=& C^{res}_{\rho\pi}  D_\phi(s)
+C^\omega_{\rho\pi} \; {\text ,}
\nn
 \enea
 where
 \bgea
 C^\omega_{\rho\pi} & = & -16\pi \alpha \; \frac{4\sqrt{2}
h_V}{9f_\pi^2}\; ( \sqrt{2} h_V - \sigma_V f_V s D_\omega(s) )
\nn \\
 \label{eq:C-omega-rhopi}
& \approx & (0.091 - 0.002\;i) \ \text{GeV}^{-2}
 \enea
 and
 \bgea C^{res}_{\rho\pi} &=&  -16\pi \alpha \; \frac{4\sqrt{2}
h_V}{9f_\pi^2}\; \sqrt{2} \; \sigma_V \; \varepsilon_{\omega\phi} \; f_V\; s
\nn\\
 \label{eq:C-res-rhopi}
 &\approx& - 0.0052.
\enea
The KLOE values for these constants are
$C^{res}_{\rho\pi}\approx - 0.0057$ and $C^\omega_{\rho\pi} =
0.26 \ \text{GeV}^{-2}$.
However, in the experiment, they are entangled
and one has to compare the total contributions.
Using the values~(\ref{eq:C-omega-rhopi})
and~(\ref{eq:C-res-rhopi}) we have
$|C_{\rho\pi}(M_\phi^2)| \approx 1.2$, which
is in a reasonable agreement with KLOE fit
$|C_{\rho\pi}(M_\phi^2)| \approx 1.3$.

\subsection{Full model prediction for the cross section}

\begin{figure}
\begin{center}
\resizebox{0.45\textwidth}{!}{%
     \includegraphics{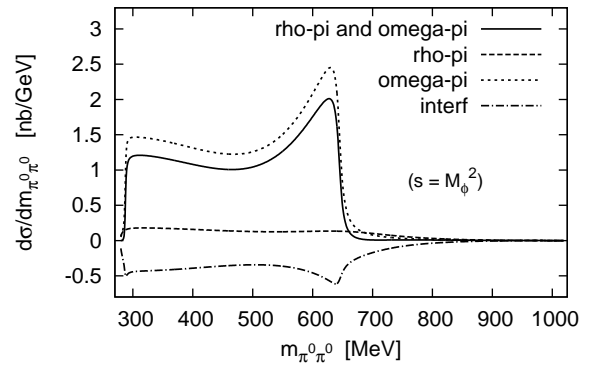}
      }
\end{center}
\caption{Vector and double-vector decay contributions to $d
\sigma/d \sqrt{p^2}$ of $e^+e^-\to \pi^0 \pi^0 \gamma$ at
$\sqrt{s}=M_\phi$ in the approximation
$\varepsilon_{\omega\phi}=0.058$, $\varepsilon^\prime=-0.0026$. The
($\phi\pi$) channel is negligible and not shown in the plot }
\label{fig:vec-interf:Mphi}
\end{figure}

Interference of leading vector resonance contributions $(\rho\pi)$
and $(\omega\pi)$ is presented in Fig.~\ref{fig:vec-interf:Mphi}.
One can see a destructive interference.

\begin{figure}
\begin{center}
\resizebox{0.45\textwidth}{!}{%
     \includegraphics{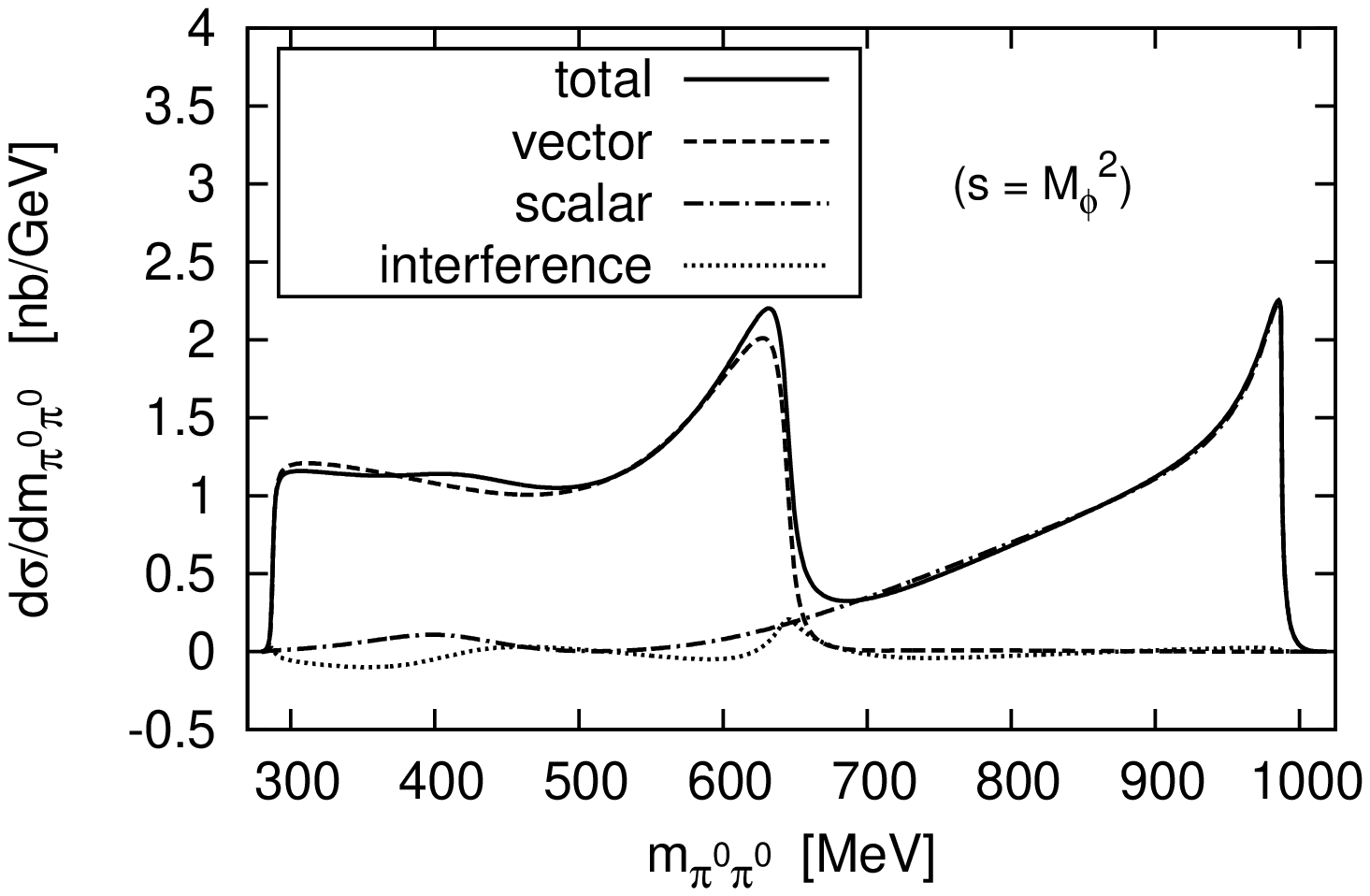}
      }

\resizebox{0.45\textwidth}{!}{%
     \includegraphics{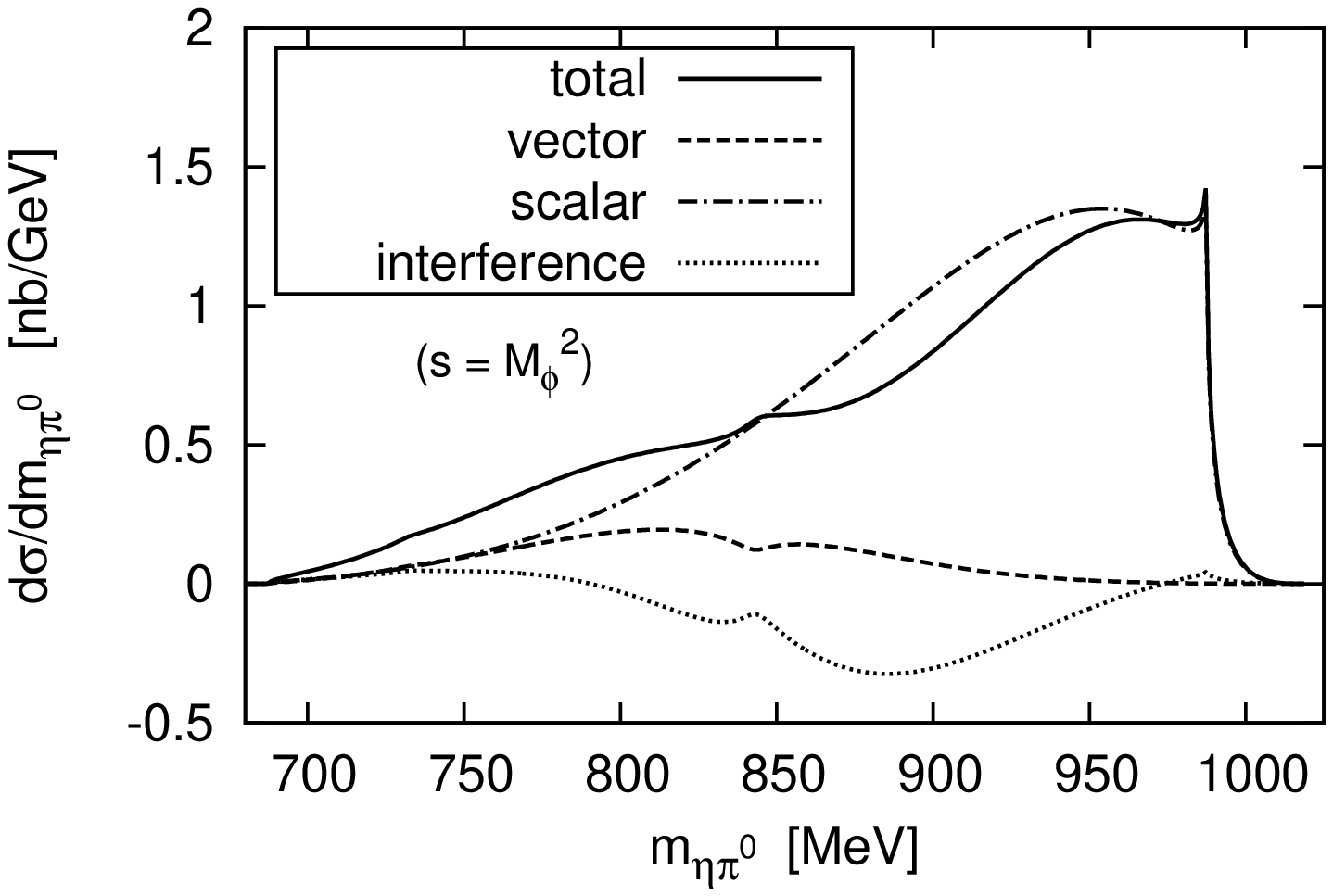}
      }
\end{center}
\caption{Differential cross section $d \sigma/d \sqrt{p^2}$
of the $e^+e^-$ annihilation
to $\pi^0 \pi^0 \gamma$ (top panel)
and $\pi^0 \eta \gamma$ (bottom panel)
for $\sqrt{s}=M_\phi$}
\label{fig:num:Mphi}
\end{figure}

The interplay of the scalar~(\ref{fsr_proc_scal})
and vector decay~(\ref{fsr_proc_vec}) contributions
to $d \sigma/d \sqrt{p^2}$ is shown in
Fig.~\ref{fig:num:Mphi} (for $\sqrt{s}=M_\phi$).
One observes a complicated interference between vector and scalar contributions.
We see that in the case of the $\pi^0\pi^0\gamma$ final state
the  vector contribution has the
same size as the scalar meson one and  is much smaller than the scalar
one for the $\pi^0\eta\gamma$ final state.

\begin{figure}
\begin{center}
\resizebox{0.45\textwidth}{!}{%
     \includegraphics{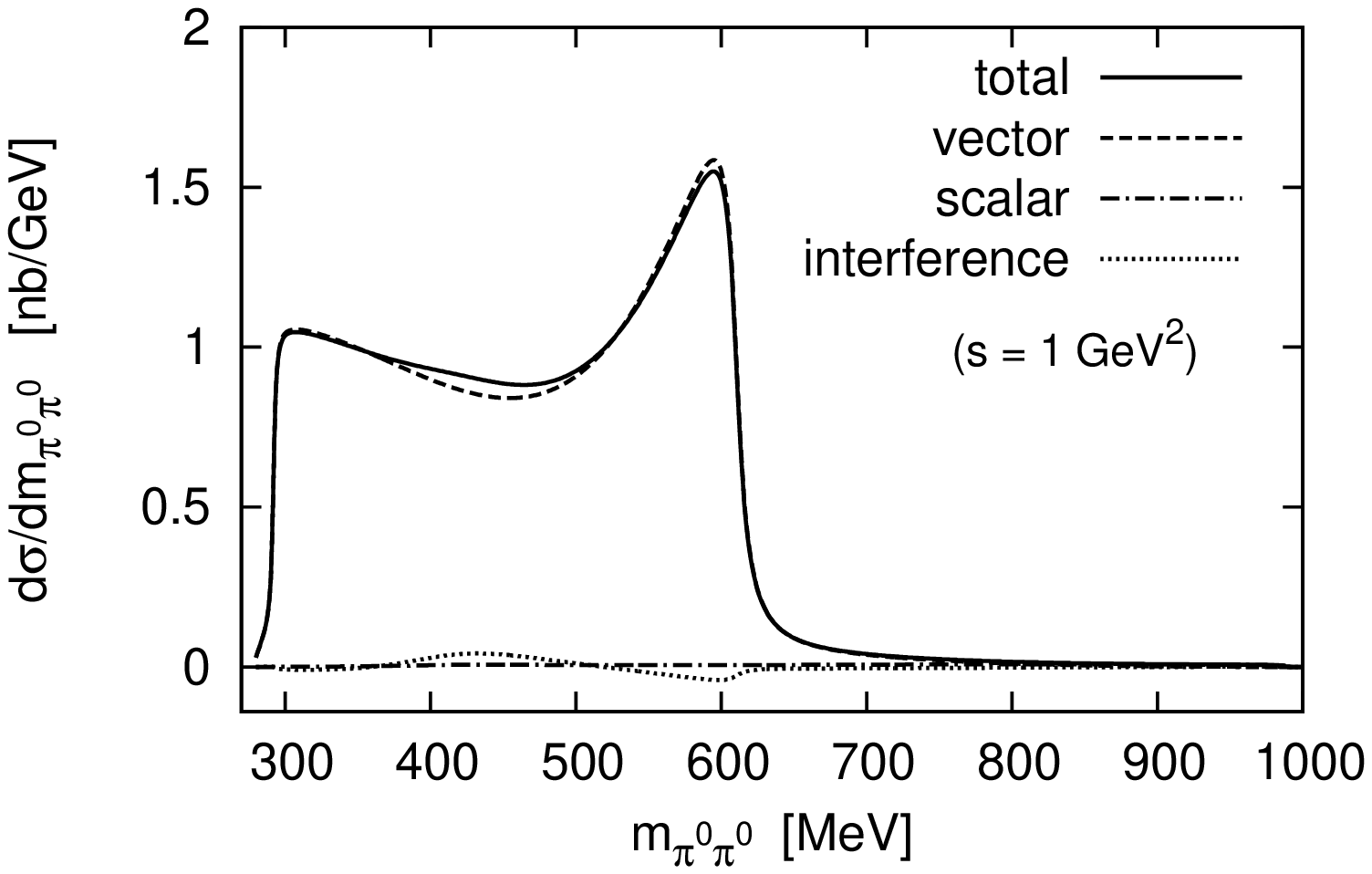}
      }

\resizebox{0.45\textwidth}{!}{%
     \includegraphics{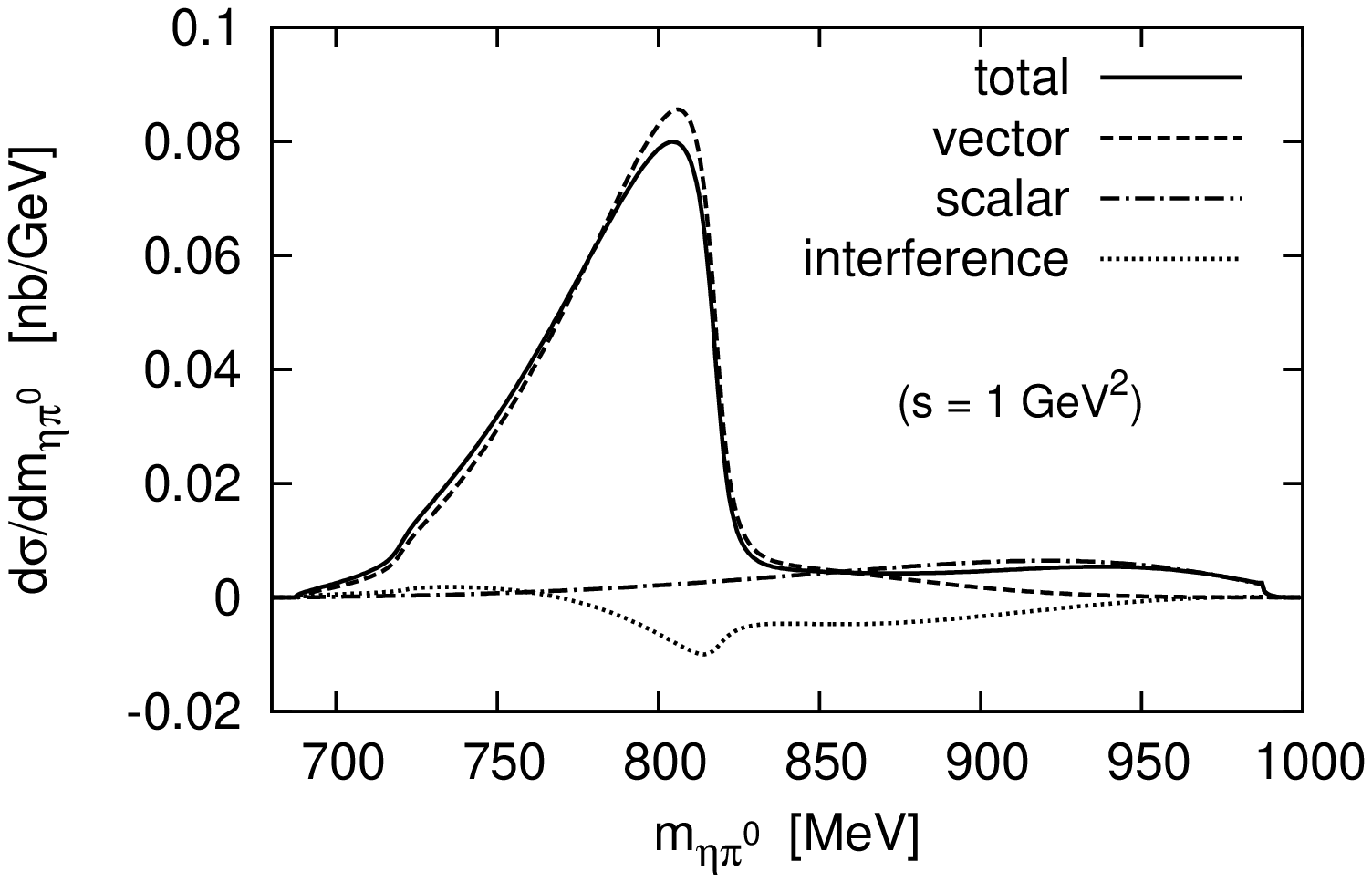}
      }
\end{center}
\caption{Differential cross section $d \sigma/d \sqrt{p^2}$
of the $e^+e^-$ annihilation
to $\pi^0 \pi^0 \gamma$ (top panel)
and $\pi^0 \eta \gamma$ (bottom panel)
for $\sqrt{s}=1$~GeV}
\label{fig:num:1GeV}
\end{figure}

%%%%%%%%%%%%%%%%%%%%%%%%%%%%%%%%%%%%%%%%%%%
% OFF PEAK
Notice that there exist the off-peak ($\sqrt{s} = 1$~GeV) data
collected by KLOE. The $\phi$ meson decays get strongly suppressed
and the total cross section is determined by the vector
contribution only.
In order to support the related activity
and provide the important model estimates, we include
this case into our numerical calculation. The corresponding results
are presented in Fig.~\ref{fig:num:1GeV}.

%%%%%%%%%%%%%%%%%%%%%%%%%%%%%%%%%%%%%%%%%%%

%=====================================================%
%                           CONCLUSIONS
%
%=====================================================
\section{Conclusions}
\label{section_conlus}

We presented a general framework for the model-independent decomposition
of the differential cross section for the
final-state radiation in the reactions
$e^+e^- \to \pi^0\pi^0\gamma$ and $e^+e^- \to \pi^0\eta \gamma$,
for which the ISR contribution is absent and the leading-order
cross section is determined solely by the FSR mechanism.

We calculated the explicit form of the functions $f_i$, which
carry the model-dependent information about the processes. 
Scalar resonance, vector and double
vector meson exchange contributions are considered. Notice that
all the relative phases are fixed from the Lagrangian of Resonance Chiral Theory.
The only exception is the sign of the $\varepsilon^\prime$ parameter, which
is related to a rare $\phi\to\omega\pi$ decay.

The Lagrangian is taken at the linear-in-resonance level in the
even-intrinsic-parity sector and at the bilinear-in-resonance level in
the odd-intrinsic-parity sector. 
For agreement with data, the \rxt Lagrangian with the lowest nonet of
vector and scalar mesons~\cite{EckerNP321} was extended by including some
$SU(3)$ symmetry breaking effects.
 At the same time, we tried to keep 
 the number of model parameters
as small as possible, using additional constraints.
The model parameters for the scalar sector were obtained from the
fit~\cite{Ivashyn:2009te} to the KLOE
data~\cite{Aloisio:2002bsa,KLOEres}.

As a by-product, we also obtained predictions for various transition
form factors: \  $\gamma^\ast \gamma S$, $SPP$, $\gamma^\ast VP$ and
$\gamma^\ast PP$. These expressions follow directly from the Lagrangian,
and the corresponding parameters are fixed to a large extent.

The numerical results for the differential cross section $d
\sigma/d \sqrt{p^2}$ are given for two cases: $\sqrt{s} = 1$~GeV
and $\sqrt{s}=M_\phi$  and demonstrate an interplay of the scalar
and vector decay contributions. The influence of the scalar and
vector contributions on the cross section is studied in detail.

The main conclusions of the numerical studies are the following:

\begin{itemize}
\item for the $\pi^0\eta\gamma$ final state the vector
contribution is much smaller than the scalar one at
$\sqrt{s}=M_\phi$ whereas for the $\pi^0\pi^0\gamma$  channel the
vector and scalar contributions are of the same size;

\item among the vector contributions to the $\pi^0\pi^0\gamma$ channel
the leading one comes from the
$\gamma^\ast(\to(\rho;\phi))\to\omega\pi$ mechanism; comparing to
the KLOE fit~\cite{KLOEres:07} we have concluded that about $85\%$
of this contribution is caused by the $\rho$ intermediate state,
and the rest can be explained either by the $\rho(1450)$ or by the
G-parity-violating process: $\gamma^\ast\to\phi\to\omega\pi$. New
experimental data at $\sqrt{s} = 1$~GeV can help to clarify which
of these two mechanisms is responsible for the rest;

\item  at $\sqrt{s} = 1$~GeV the scalar contribution is
suppressed and the total cross section is determined only
by the vector contribution both for the
$\pi^0\pi^0\gamma$ and $\pi^0\eta\gamma$ channels.
\end{itemize}

At the end, we would like to emphasize that the developed approach
allows one to obtain the cross section and branching fraction
close to the experimental results. The main advantage of this
approach is a small number of model parameters.

The proposed framework can be implemented in a Monte Carlo generator,
for the inspection of the completely differential characteristics of
the reaction, and thus is useful for a data analysis and a detailed comparison
of various models.

%=====================================================
%\section*{Acknowledgments}

% \vspace{5pt}
{
\acknowledgement
 \noindent{\it Acknowledgements. }
We would like to thank
Zurab Silagadze for
his comments on~\cite{Achasov:1999wr}
and for providing us with a copy
of~\cite{EidelmanKuraev}.
This paper profited from discussions with
Henryk Czy\.z.
S.E., A.K. and O.S. acknowledge partial support by the INTAS grant 05-1000008-8328
``Higher order effects in $e^+ e^-$ annihilation and muon
anomalous magnetic moment''.
S.E.~acknowledges partial support by RFFI grant 09-02-01143.
S.I.~was supported in part by Research Training Network EU-MRTN-CT-2006-035482
(FLAVIAnet).
G.P.~is grateful to the MIT Center of Theoretical Physics for hospitality
 while this work was being written.
\endacknowledgement
}

%=====================================================%
%                           APPENDIX
%
%=====================================================
\appendix{}

 \def\theequation{\Alph{section}\arabic{equation}}
 \setcounter{section}{0}
 \def\appendixprefix{Appendix~}
 \def\thesection{\appendixprefix\Alph{section}}

\newcommand\appsection[1]{
  \section[]{\!:~#1}
  \setcounter{equation}{0}
  \setcounter{subsection}{0}
  }

%=====================================================
\appsection{Pseudoscalar mesons, scalar multiplet and
         the \rxt Lagrangian
         }
\label{App:A}

In chiral theory, the pseudoscalar mesons $\pi$, $K$, $\eta$ can be treated as
pseudo-Nambu-Goldstone bosons of spontaneous
$G=SU(3)_L\times SU(3)_R$ to $H=SU(3)_V$ broken symmetry.
The physical states $\eta$, $\eta^\prime$ can be introduced using the scheme
with two mixing angles ($\theta_0$, $\theta_8$), for a review
see~\cite{Feldmann:rev}.
The adopted scheme is consistent with
chiral theory and takes into account the effects of
$U(1)$ axial anomaly  and $SU(3)$ flavor breaking ($m_s \gg m_{u,d} $).
In our notation~\cite{Ivashyn:2009te} the pseudoscalar nonet reads
\bgea
\label{eq:ps-nonet}
u &= &
 \\ \nn
 &\exp& \left\{ \frac{i}{\sqrt{2} f_\pi}
\left(
\begin{array}{ccc}
 \frac{\pi^0 + C_{q}\eta + C_{q}^\prime \eta^\prime}{\sqrt{2}} & \pi^+  & \frac{f_\pi}{f_K} {K^+ } \\
 \pi^- & \hspace{-0.18in} \frac{-\pi^0+ C_{q}\eta + C_{q}^\prime \eta^\prime}{\sqrt{2}} & \frac{f_\pi}{f_K} {K^0} \\
 \frac{f_\pi}{f_K} {K^-} & \frac{f_\pi}{f_K} {\bar{K}^0} & \hspace{-0.06in} - C_{s}\eta + C_{s}^\prime \eta^\prime
\end{array}
\right)
\right\}
,
\enea
where
\bgea
\label{eq:eta-coefficients}
C_q &\equiv& \frac{f_\pi}{\sqrt{3} \cos(\theta_8-\theta_0)}
\left( \frac{1}{f_0}\cos\theta_0 - \frac{1}{f_8}\sqrt{2} \sin\theta_8 \right),
\\
C^\prime_q &\equiv& \frac{f_\pi}{\sqrt{3} \cos(\theta_8-\theta_0)}
\left(\frac{1}{f_8} \sqrt{2} \cos\theta_8 + \frac{1}{f_0}\sin\theta_0 \right),
\nn
\\
C_s &\equiv& \frac{f_\pi}{\sqrt{3} \cos(\theta_8-\theta_0)}
\left( \frac{1}{f_0}\sqrt{2}\cos\theta_0 +  \frac{1}{f_8}\sin\theta_8 \right),
\nn
\\
C^\prime_s &\equiv& \frac{f_\pi}{\sqrt{3} \cos(\theta_8-\theta_0)}
\left(\frac{1}{f_8}\cos\theta_8 -  \frac{1}{f_0}\sqrt{2}  \sin\theta_0 \right)
\nn
.
\enea

The vielbein field which represents the pseudoscalar mesons is
$u_\mu = \imnim u^+ D_\mu u^+$ and
$\chi_+ = u^+ \chi u^+ + u\chi u$ is the explicit symmetry-breaking term,
$\chi \approx \mathrm{diag}(m_\pi^2,\ m_\pi^2,\ 2 m_K^2-m_\pi^2)$
in the isospin symmetry limit.

The electromagnetic field $B^\mu$
is included as an external source, $F_{\mu\nu}=\partial_\mu B_\nu -
\partial_\nu B_\mu$ is the electromagnetic field tensor.
It appears in the chiral covariant derivative,
which in our case is reduced to $D_\mu U = \partial_\mu U + \imnim e B_\mu[U,Q]$
and in the tensor $f_{+}^{\mu\nu} = e F^{\mu\nu} (u Q u^{+} + u^{+} Q u)$,
where the quark charge matrix $Q =\mathrm{diag}({2}/{3},-{1}/{3},-{1}/{3})$.

For calculations in the even-intrinsic-parity sector,
the leading-order \rxt Lagrangian for
pseudoscalar, scalar, vector mesons and
photons was derived by Ecker~et~al.~\cite{EckerNP321}.
The spin-$1$ mesons are described by
antisymmetric matrix tensor fields $V^{\nu \mu}$
and this Lagrangian is equivalent to the ChPT Lagrangian
at order $\mathcal{O}(p^{4})$
(see~\cite{EckerNP321,EckerPLB223} for details).
In our application we have somewhat released the
rigor of \rxt and use
different masses of resonances ($M_\rho \neq M_\omega \neq M_\phi$
and $M_\sigma \neq M_{a_0} \neq M_{f_0}$)
without specifying a pattern of flavor symmetry breaking
(cf. Ref.~\cite{Cirigliano:resonances}).

Interaction terms for the pseudoscalar and vector mesons read
\begin{eqnarray}
\label{lagr:vec:master}
\mathcal{L}_{vector} &=&\frac{f^{2}}{4}\left\langle u_\mu u^\mu + \chi_+ \right\rangle
\nonumber \\
&& +\frac{F_{V}}{2\sqrt{2}}\left\langle V_{\mu \nu }f_{+}^{\mu \nu
}\right\rangle +\frac{ iG_{V}}{\sqrt{2}}\left\langle V_{\mu \nu }u^{\mu }u^{\nu
}\right\rangle
,
\end{eqnarray}
here $\left\langle \cdots \right\rangle$ stands for the trace in flavor
space.

For scalar mesons we assume
the nonet symmetry of the interaction terms and
multiplet decomposition
\bgea \label{eq:multiplet_sc}
\left\{
\begin{aligned}
a_0 =& S_3
,\\
f_0 =& S_0\, \cos \theta - S_8\, \sin \theta
,\\
\sigma =& S_0\, \sin \theta + S_8\, \cos \theta ,
\end{aligned}
\right.
 \enea
where $S_3$ is the neutral isospin-one, $S_8$ is the isospin-zero
member of the flavor octet. The angle $\theta$
is the octet-singlet mixing parameter, and $\sigma \equiv f_0(600)$.
The interaction Lagrangian for scalars takes the form
\begin{eqnarray}
\label{lagr:master}
\mathcal{L}_{scalar} &=& c_d
\left\langle S u_\mu u^\mu \right\rangle + c_m \left\langle S
\chi_+ \right\rangle
.
\end{eqnarray}
There are known problems with a rigorous
inclusion of $\sigma$ and $f_0(980)$ into any \rxt
multiplet~\cite{Cirigliano:resonances}.
However, there is also a number of successful
applications~\cite{Hooft:2008we,Black:1998wt} of a scheme similar to~(\ref{eq:multiplet_sc}).
In studies of $\phi$ radiative decays this scheme was also applied
in~\cite{Ivashyn:2009te,Ivashyn:2009yt}.

Due to nonet symmetry,
the relation for scalar singlet $S_0$ and octet $S^{oct}$ coupling constants
holds,
$c_{m,d} S = c_{m,d}\left( S^{oct} + S_0/\sqrt{3} \right)$
.
In nomenclature of Ref.~\cite{EckerNP321}
this relation implies $\tilde{c}_{m,d} = c_{m,d}/\sqrt{3}$.

In the odd-intrinsic-parity sector 
the flavor $SU(3)$ symmetric Lagrangian~\cite{EckerPLB237,Prades}
reads
\bgea
\label{lagr:odd-master}
\mathcal{L}_{odd} &=&
\;\;\;\; h_V \epsilon_{\mu\nu\alpha\beta}
\left\langle V^\mu ( u^\nu f_+^{\alpha\beta} + f_+^{\alpha\beta}u^\nu )  \right\rangle
\\
&&+ \; \sigma_V \epsilon_{\mu\nu\alpha\beta} \left\langle V^\mu (
u^\nu \partial^\alpha V^\beta + \partial^\alpha V^\beta u^\nu )
\right\rangle \nn. \enea

%=====================================================%
%                          Appendix B
%
%=====================================================
\appsection{Model parameters}
\label{App_B}
\subsubsection*{Masses}
The following values for the meson masses are used in our
numerical calculations~\cite{PDG_2008}: $M_\rho = 775.49$~MeV,
$M_\omega = 782.65$~MeV, $M_\phi = 1019.456$~MeV, $m_\pi =
m_{\pi^\pm} = 139.57$~MeV, $m_{\pi^0} = 134.98$~MeV, $m_K =
493.68$~MeV, $m_\eta = 547.75$~MeV.

\subsubsection*{Mixing parameters}
The values of the $\eta$ mixing angles $\theta_0 = -9.2^\circ \pm 1.7^\circ$ and $\theta_8 = -21.2^\circ \pm 1.6^\circ$
are used~\cite{Feldmann:98},
thus $f_8= (1.26 \pm 0.04) f_\pi$ and $f_0=(1.17 \pm 0.03) f_\pi$,
where $f_\pi \approx 92.4$~MeV.
Thus, one obtains $C_q \approx 0.738$ and
$C_s \approx 0.535$.

The $\omega\phi$ mixing is given by one parameter
$\varepsilon_{\omega\phi}=0.058$~\cite{Dolinsky:1991vq}.
The states of ``ideal mixing'' $\omega_{id}=(u \bar{u} + d \bar{d})/ \sqrt{2}$ and
$\phi_{id}= s \bar{s}$ are expressed in terms of the physical ones
(mass eigenstates) as
 \begin{eqnarray}
 \label{eq:omega-phi-mix}
\omega_{id} & = & \omega + \varepsilon_{\omega\phi} \phi , \\
\phi_{id}   & = & \phi   - \varepsilon_{\omega\phi} \omega
.\nonumber
 \end{eqnarray}
In order to include a G-parity-violating $\phi\omega\pi^0$ vertex
we determine the parameter $\varepsilon^\prime$ from the
$\phi\to\omega\pi$ decay width: \bge
\Gamma(\phi\to\omega\pi)=\frac{|g_{\phi\omega\pi}|^2
p_\pi^3}{12\pi}, \ene
 where $p_\pi = \sqrt{(M_\phi^2 + m_\pi^2 -
M_\omega^2)^2/(4\; M_\phi^2) - m_\pi^2}$, the effective coupling
in our formalism is
$g_{\phi\omega\pi}=4\sigma_V\varepsilon^\prime/f_\pi$. Using the
experimental value for the $\phi\to\omega\pi$ decay branching
ratio
%$B = (4.8 ^{+1.9}_{-1.7} \pm 0.8)\times 10^{-5}$~\cite{Achasov:1999jc}
$B = (4.4 \pm 0.6)\times 10^{-5}$~\cite{Ambrosino:2008gb} and
$\sigma_V = 0.34$ one obtains $|\varepsilon^\prime|= 0.0026$.

\subsubsection*{Couplings in the even-intrinsic-parity sector}

The condition $ F_V = 2\ G_V$ for the model couplings
is used in our calculation  to make the one-loop
amplitude finite~\cite{Ivashyn:2007yy} without use of counter-terms.
This relation has been addressed in~\cite{EckerPLB223}
in a different context, namely it has been shown that
the constraints imposed by the high-energy behavior of the vector and
axial-vector FF's lead to it, in addition to the
relation $F_V G_V = f_\pi^2$.
Note that $ F_V = 2\ G_V$ also appears
in alternative models, e.g., Hidden Local Gauge Symmetry
Model and massive Yang-Mills theory for vector mesons,
see a discussion in~\cite{EckerPLB223}.
For numerical calculations we use
$G_V = f_\pi/\sqrt{2} = 65.34$~MeV,
$F_V = 2\ G_V = 130.68$~MeV.

Alternatively, respecting phenomenology,
one may fix $F_{V}$ and $G_{V}$
by means of fitting the measured partial decay widths of the vector
mesons (see, e.g.,~\cite{EckerNP321}) at tree level.
In particular, for $\rho\to e^+ e^-$ one has
 \bge
 \label{eq:C2}
 \Gamma_{\rho \to e^+e^-} = \frac{e^4 F_V^2}{12 \pi M_\rho}
 \ene
and for the $\rho \to \pi \pi$ the tree level width is given by
 \bgea
\Gamma_{\rho \to \pi^+\pi^-} = \frac{G_V^2}{48 \pi f_\pi^4 }
\bigl(m_\rho^2 - 4 m_\pi^2 \bigr)^{3/2} \label{eq:C111}.
 \enea
The experimental data are the following~\cite{PDG_2008}:
 $\Gamma(\rho\to\pi^+ \pi^-) = 146.2\pm 0.7$~MeV
and $\Gamma_{\rho \to e^+ e^-} = 7.04 \pm 0.06$~keV.
Values obtained in this way
are $G_V = 65.14 \pm 0.16$~MeV and $F_V = 156.41 \pm 0.67$~MeV.
The estimated values support the $F_V \approx 2\ G_V$ conjecture.

\subsubsection*{ Couplings in the odd-intrinsic-parity sector}

The coupling constant $f_V$ is given by $f_V = {F_V}/{M_\rho} \approx0.17$.
The parameter $h_V$ can be fixed from the $V\to P\gamma$ decay width,
in particular, the
$\rho \to \pi \gamma$ width
\begin{equation}
\label{eq:rhopigamma:width}
\Gamma(\rho \to \pi \gamma) = \frac{4 \alpha M_\rho^3 h_V^2 }{27 f_\pi^2
}\left(1- \frac{m_\pi^2}{M_\rho^2} \right)^3
\end{equation}
leads to $h_V = 0.041 \pm 0.003$.

One can use a special short-distance constraint of \rxt
in order to relate $\sigma_V$ to $f_V$ and $h_V$.
Namely, one can require the form factors~(\ref{eq:FFs_SU3_eta})
to vanish at $Q^2 \to - \infty$ as expected from QCD.
In this connection we refer to~\cite{Pich_IJMP,Knecht},
where in the framework of \rxt
high-energy behavior of three-point Green functions $VVP$, $VAP$, $AAP$
has been studied.

At $Q^2 \to -\infty $
the propagators~(\ref{vector-propagator-simple})
$D_V (Q^2) \to 1/ Q^2$ and we obtain the following relation (neglecting mixing)
\begin{equation}
\label{eq:qcdsum:1}
 \sqrt{2} h_V - \sigma_V f_V = 0 .
 \end{equation}
This constraint reduces the number of independent parameters in the model,
in particular, expresses the poorly known parameter $\sigma_V$ via $h_V$ and
$f_V$, which can be fixed from data.
Thus we obtain $\sigma_V \approx 0.34$.

Notice, an additional constraint on the parameters
$\sigma_V$, $h_V$ and $f_V$ follows from the
short-distance behavior of the $\gamma^\ast \gamma^\ast \pi^0$
form factor (see a discussion in Ref.~\cite{Knecht}):
\begin{equation}
\label{eq:qcdsum:2}
 -\frac{N_c}{ 4 \pi^2} + 16 \sqrt{2} h_V f_V - 8 \sigma_V f_V^2 = 0
 .
\end{equation}
It allows to  further reduce  the number of independent parameters.
For example, one can leave $f_V$ to be the only independent
parameter and deduce
from~(\ref{eq:qcdsum:1}) and~(\ref{eq:qcdsum:2})
\bgea
 \sigma_V &=& \frac{N_c}{32\; \pi^2\; f_V^2},
\nn\\
 h_V &=& \frac{N_c}{32 \sqrt{2}\; \pi^2\; f_V},
\enea
which results in the numerical values $\sigma_V = 0.329$ and $h_V = 0.0395$
--- fairly close to those obtained
with the use of Eq.~(\ref{eq:rhopigamma:width}).

In favor of broken flavor $SU(3)$
symmetry, one may introduce separate couplings
for each vector meson, i.e. replace $f_V$ by $f_{\rho}$, $f_{\omega}$, $f_{\phi}$,
and further $h_V$ by $h_{\gamma \rho \pi}$, $h_{\gamma \omega \pi}$,
$h_{\gamma \rho \eta},\ldots$
($\gamma VP$ transition)
, and also $\sigma_V$ by
$\sigma_{\omega \rho \pi}$, $\sigma_{\rho \rho \eta},\ldots$

\subsubsection*{Parameters for scalar mesons}

The widths for $a_0\to\gamma\gamma$ and $f_0\to\gamma\gamma$  decays
are expressed in terms of~(\ref{scalar-two-photon-ff-1})--(\ref{scalar-two-photon-ff-3}), %-(\ref{scalar-two-photon-ff-3}),
for example:
 \bgea
 \label{width:agg}
 \Gamma_{a_0\to\gamma\gamma} &=& \frac{e^4 p^4}{64\pi \sqrt{p^2}}
 |G_{a_0\gamma\gamma}^{(K)}(p^2,0)|^2
 .
 \enea
The strong decay widths of
the scalar mesons in the lowest order (tree level) are
 \bgea \label{width:ape}
\Gamma_{a_0\to\pi\eta}(p^2) &=& \!\frac{\!|G_{a_0\!\pi \eta}(p^2)|^2\!\!}{8 \pi p^2}
\nn\\&&\times
\sqrt{\!\frac{(p^2\!+\!m_\pi^2\!-\!m_\eta^2)^2}{4p^2}-\!m_\pi^2},
\\
%\label{width:fpp}
\Gamma_{f_0\to \pi\pi}(p^2) &=& (1+\frac{1}{2})
\frac{|G_{f_0\pi
\pi}(p^2)|^2}{8 \pi p^2}
\nn\\&&\times
\sqrt{p^2/4 - m_\pi^2},
\nn\\
\nn%\label{width:akk}
  \!\!  \Gamma_{a_0\to K\bar{K}}(p^2) &=& 2 \frac{|G_{a_0KK}(p^2)|^2}{8 \pi p^2}
  \sqrt{p^2/4 - m_K^2}
 , \\
\nn%\label{width:fkk}
  \!\!  \Gamma_{f_0\to K\bar{K}}(p^2) &=& 2 \frac{|G_{f_0KK}(p^2)|^2}{8 \pi p^2}
  \sqrt{p^2/4 - m_K^2}
  ,
 \enea
where ${p^2}$ is the invariant mass squared of the scalar meson;
see also definition~(\ref{eq:SPP-ffs:1}),~(\ref{eq:SPP-ffs:2}).
For discussion of momentum-dependent couplings $G_{SPP}(p^2)$ and
constant $SPP$ couplings of other models (e.g.,~\cite{Achasov_Ivanchenko})
see Ref.~\cite{Ivashyn:2009yt}.

The finite-width effects for scalar resonances are very important
and expressions~(\ref{width:agg}), (\ref{width:ape}) do not have
physical meaning of decay width, when evaluated at the
resonance peak value of $p^2$.
Nevertheless, in several papers, e.g.,~\cite{Ivashyn:2007yy,EckerNP321,Guo:2009hi},
the tree-level expressions of a similar form
 were used to find the
 model parameters ($c_d$, $c_m$ and $\theta$) from
 measured widths.
It was observed~\cite{Ivashyn:2009te,Ivashyn:2009yt,Ivashyn:2008sg}
that the coupling constants could be better determined from fitting the
$\pi\pi$ and $\pi \eta$ invariant mass distributions in
$e^+ e^- \to \phi \to\gamma\pi\pi$ and $e^+ e^- \to \phi \to \gamma\pi \eta$
reactions.
The fit results~\cite{Ivashyn:2009te} are shown in Table~\ref{tab:fit-results}
and these values are used in our numerical calculations.
Notice that for this fit we used data from~\cite{KLOEres}
($\pi^0\pi^0\gamma$) and~\cite{Aloisio:2002bsa}
($\pi^0\eta\gamma$). Recently, a new KLOE result
for the latter appeared~\cite{Ambrosino:2009py}, and we find
reasonable agreement with it without refitting, see a discussion
in Section~\ref{section_numer}.

\begin{table}
\caption{Scalar meson parameters~\cite{Ivashyn:2009te}.
Couplings and mass parameters are given in MeV.
}
\label{tab:fit-results}
\begin{center}
\small
\begin{tabular}{|cc|ccc|c|}
\hline
$c_d$ & $c_m$ & $M_{a_0}$ & $M_{f_0}$ &$M_{\sigma}$  & $\theta$
\\
\hline
$93^{+11}_{-5}$ & $46^{+9}_{-2}$ & $1150^{+50}_{-23}$ & $986.1^{+0.4}_{-0.5}$ & $504^{+242}_{-53}$  & $36^o\pm2^o$
\\
\hline
\end{tabular}
\end{center}
\end{table}

%=====================================================%
%                          Appendix C
%
%=====================================================
\appsection{Example of the factorization of the
$\gamma^\ast \to \gamma f_0\to \gamma \pi^0\pi^0$ transition amplitude}
\label{App_C}

In this Appendix we sketch
the general structure of the scalar meson contribution
$f_1^{S}$ giving emphasis on the appearance of
the electromagnetic form factors of the pseudoscalars
in the formulae~(\ref{f1-scalar-f0}) and~(\ref{f1-scalar-a0}).

Consider the part of the $M^{\mu\nu}$ amplitude~(\ref{eqn:fsr}) of $\gamma^\ast \to \gamma
f_0\to \gamma \pi^0\pi^0$ with a pion loop transition, $M^{\mu\nu}_{\pi\ loop}$.
Figure~\ref{fig:g-g-S-scheme} is of help and
one observes two terms
\bgea
 M^{\mu\nu}_{\pi\ loop}
 &=&
M^{\mu\nu}_{\gamma \to \pi\ loop} + M^{\mu\nu}_{\gamma \to V \to \pi\ loop},
 \enea
 the former with the contact $\gamma^\ast \to \pi^+\pi^-$ coupling
%($M^{\mu\nu}_{\gamma \to \pi\ loop}$)
and the latter with an intermediate vector resonance.
%($M^{\mu\nu}_{\gamma \to V \to \pi\ loop}$).
They read
 \bgea
\label{m:gamma-TO-pi} M^{\mu\nu}_{\gamma \to \pi loop}
 &=&
 \frac{-4 e^2 \imnim}{(4\pi)^2}\tau_1^{\mu\nu} G_{f_0\pi\pi}(p^2)
 \frac{2}{m_\pi^2}  I\left(\frac{Q^2}{m_\pi^2}, \frac{p^2}{m_\pi^2}\right)
 \nn\\&&\times
 D_{f_0}(p^2)G_{f_0\pi\pi}(p^2)
\\
\label{m:gamma-TO-Vpi}
M^{\mu\nu}_{\gamma \to V \to \pi loop}
 &=&
 \frac{-4 e^2 \imnim}{(4\pi)^2}\tau_1^{\mu\nu} G_{f_0\pi\pi}(p^2)
 \frac{2}{m_\pi^2}  I\left(\frac{Q^2}{m_\pi^2}, \frac{p^2}{m_\pi^2}\right)
 \nn\\&&\times
 D_{f_0}(p^2)G_{f_0\pi\pi}(p^2)
\nn\\
&&\times\; \frac{1}{f_\pi^2} F_V G_V Q^2 D_\rho(Q^2).
 \enea
The $\rho$ meson propagator $D_\rho(Q^2)$ is given
in~(\ref{vector-propagator-simple}).
The form factor
$G_{f_0\pi\pi}(p^2)$ is given by~(\ref{eq:SPP-ffs:1}).
The loop integral $I(a,b)$ can be found, e.g.,
in~\cite{Close:1992ay} and~\cite{Bramon:2002iw}, and reads
\bgea
I(a,b) &=& \frac{1}{2(a-b)}\ -\ \frac{2}{(a-b)^2}
           \left[f\left(\frac{1}{b}\right) - f\left(\frac{1}{a}\right) \right]
       \nn\\
       &+& \frac{a}{(a-b)^2}
           \left[g\left(\frac{1}{b}\right) - g\left(\frac{1}{a}\right) \right]
       ,
\enea
with
\bgea
f(x) &=&
   \left\{
   \begin{array}{lr}
      - \left[\arcsin\left(\frac{1}{2\sqrt{x}}\right)\right]^2, & x > \frac{1}{4},
      \\
      \frac{1}{4} \left[\log\frac{n_+(x)}{n_-(x)} - \imnim \pi \right]^2, & 0 < x < \frac{1}{4},
      \\
      \left[\log\frac{1+\sqrt{1-4x}}{2\sqrt{-x}}\right]^2, & x < 0,
   \end{array}
   \right.
\nn
\\
g(x) &=&
   \left\{
   \begin{array}{lr}
      \sqrt{4x - 1}\ \arcsin\left(\frac{1}{2\sqrt{x}}\right), & x > \frac{1}{4},
      \\
      \frac{1}{2}\sqrt{1-4x} \left[\log\frac{n_+(x)}{n_-(x)} - \imnim \pi \right], & 0 < x < \frac{1}{4},
      \\
      \sqrt{1-4x}\ \log\frac{1+\sqrt{1-4x}}{2\sqrt{-x}}, & x < 0,
   \end{array}
   \right.
\nn
\\
n_{\pm}(x) &=& \frac{1}{2x} \left( 1 \pm \sqrt{1 - 4x} \right)
\enea
For a reference, we remind the alternative notation of~\cite{Ivashyn:2007yy}:
 \bgea
 \Psi(m^2, p^2, Q^2) &=& (a-b)I(a,b),
 \nn\\
 {1}/({Q\cdot k}) &=& {2}/({Q^2 - p^2}),
 \enea
with $a=Q^2/m^2$ and $b=p^2/m^2$.

Combining~(\ref{m:gamma-TO-pi}) and~(\ref{m:gamma-TO-Vpi}),
one finds
\bgea
\label{m:gamma-TO-pi-total}
 M^{\mu\nu}_{\pi loop}
&=&
 \frac{-4 e^2 \imnim}{(4\pi)^2}\tau_1^{\mu\nu} G_{f_0\pi\pi}(p^2)
 \frac{2}{m_\pi^2}  I\left( \frac{Q^2}{m_\pi^2},  \frac{p^2}{m_\pi^2}\right)
  \nn\\\nn&&\times
 D_{f_0}(p^2)G_{f_0\pi\pi}(p^2)\,F_{em}^{\pi}(Q^2)
\\\nn
&\equiv& - \imnim e^2 \tau_1^{\mu\nu} 
D_{f_0}(p^2)G_{f_0\pi\pi}(p^2)
G_{f_0\gamma^\ast\gamma}^{(\pi)}(p^2,Q^2) ,
\enea
where
the two-photon form factor of a scalar meson $G_{f_0\gamma^\ast\gamma}^{(\pi)}(p^2,Q^2)$ is
given in~(\ref{scalar-two-photon-ff-1}).
The pion electromagnetic form factor $F_{em}^{\pi}(Q^2)$ in \rxt
is given by
 \bgea
 \label{pion-em-ff}
 F_{em}^{\pi} (Q^2) &=& 1 - \frac{F_V G_V}{f_\pi^2} Q^2 D_\rho(Q^2).
\enea

\begin{figure}
\begin{center}
\resizebox{0.48\textwidth}{!}{%
   \includegraphics{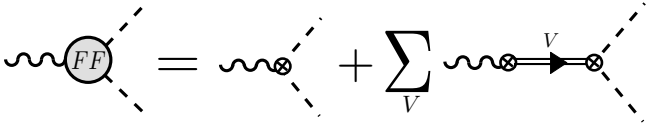}
   }
\end{center}
\caption{The $\mathcal{O}(p^2)$ electromagnetic vertex of a
(off-mass-shell) pseudoscalar meson in
\rxt. All possible intermediate vector resonances $V =\rho^0, \omega, \phi, ...$ in
general contribute.
For real photons only the first term on the r.h.s. is non-zero. }
\label{fig:FFScheme}
\end{figure}

%---------------------------
Factorization in the part of the $M^{\mu\nu}$ amplitude~(\ref{eqn:fsr})
 with a kaon loop transition, $M^{\mu\nu}_{K\ loop}$, is analogous.
The kaon form factor in \rxt is
 \bgea
\label{kaon-em-ff}
 F_{em}^{K} (Q^2) &=& 1 - \frac{F_V G_V}{f_K^2} Q^2
\left( \frac{1}{2} D_\rho(Q^2) \right.
\nn\\
&&
\left. + \frac{1}{6} D_\omega(Q^2) + \frac{1}{3}
D_\phi(Q^2) \right) ,
 \enea
The vector meson $V= \rho, \omega, \phi$  propagators are
given by~(\ref{vector-propagator-simple}).
The form factors in form~(\ref{pion-em-ff}) and~(\ref{kaon-em-ff})
include contributions from the photon--vector transition (vector
meson dominance, VMD) and the direct $\gamma P P$ interaction, see
Fig.~\ref{fig:FFScheme}.
The detailed discussion of two versions of
VMD (VMD1 and VMD2) is given in the review~\cite{O'Connell}.
It turns out that the \rxt corresponds
to the VMD1 version.

For discussion of the one-loop modification of the electromagnetic vertex
and \rxt-motivated calculation of the kaon form factor
see~\cite{Ivashyn:2006gf}.

%=====================================================
%
%                           BIBLIOGRAPHY
%
%=====================================================

%=====================================================
\end{document}